\documentclass{JINST}
\let\ifpdf\relax
\usepackage{amsmath}
\usepackage{graphicx}
\usepackage{subfig}
\usepackage{multirow}
\usepackage{hepparticles}

\newcommand{\Cf}{$^{252}$Cf }
\newcommand{\Bi}{$^{207}$Bi }

\newcommand{\nuebar}{\HepAntiParticle{\nu}{e}{} }
\newcommand{\betaplus}{\HepGenParticle{\beta}{}{+} }

\newcommand{\nistAck}{Certain trade names and company products are mentioned in the text or identified in illustrations in order to adequately specify the experimental procedure and equipment used. In no case does such identification imply recommendation or endorsement by the National Institute of Standards and Technology, nor does it imply that the products are necessarily the best available for the purpose. }

\graphicspath{ {./Figures/}}

\title{Light Collection and Pulse-Shape Discrimination in Elongated Scintillator Cells for the PROSPECT Reactor Antineutrino Experiment} \author{J.~Ashenfelter$^a$, B.~Balantekin$^b$, H.~R.~Band$^a$, G.~Barclay$^c$, C.~D.~Bass$^d$, D.~Berish$^e$, N.~S.~Bowden$^f$, A.~Bowes$^g$, J.~P.~Brodsky$^f$, C.~D.~Bryan$^c$, J.~J.~Cherwinka$^h$, R.~Chu$^{i,j}$, T.~Classen$^f$, K.~Commeford$^k$, D.~Davee$^l$, D.~Dean$^i$, G.~Deichert$^c$, M.~V.~Diwan$^m$, M.~J.~Dolinski$^k$, J.~Dolph$^m$, D.~A.~Dwyer$^n$, J.~K.~Gaison$^a$, A.~Galindo-Uribarri$^{i,j}$, K.~Gilje$^g$, A.~Glenn$^f$, B.~W.~Goddard$^k$, M.~Green$^i$, K.~Han$^a$\thanks{Corresponding author.}, S.~Hans$^o$, K.~M.~Heeger$^a$, B.~Heffron$^{i,j}$, D.~E.~Jaffe$^m$, T.~J.~Langford$^a$\thanks{Corresponding author.}, B.~R.~Littlejohn$^g$\thanks{Corresponding author.}, D.~A.~Martinez~Caicedo$^g$, R.~D.~McKeown$^l$, M.~P.~Mendenhall$^p$, P.~Mueller$^i$, H.~P.~Mumm$^p$, J.~Napolitano$^e$, R.~Neilson$^k$, D.~Norcini$^a$\thanks{Corresponding author.}, D.~Pushin$^q$, X.~Qian$^m$, E.~Romero$^{i,j}$, R.~Rosero$^o$, L.~Saldana$^a$, B.~S.~Seilhan$^f$, R.~Sharma$^m$, S.~Sheets$^f$, N.~T~Stemen$^a$, P.~T.~Surukuchi$^g$,  R.~L.~Varner$^i$, B.~Viren$^m$, W.~Wang$^i$, B.~White$^i$, C.~White$^g$, J.~Wilhelmi$^e$, C.~Williams$^i$, T.~Wise$^a$, H.~Yao$^l$, M.~Yeh$^o$, Y.-R.~Yen$^k$, G.~Zangakis$^e$, C.~Zhang$^m$, X.~Zhang$^g$, (The PROSPECT Collaboration) \\
  \llap{$^a$}Wright Laboratory, Department of Physics, Yale University\\
  New Haven, CT, USA\\
  \llap{$^b$}Department of Physics, University of Wisconsin, Madison\\
  Madison, WI, USA\\
  \llap{$^c$}High Flux Isotope Reactor, Oak Ridge National Laboratory\\
  Oak Ridge, TN, USA \\
  \llap{$^d$}Department of Chemistry and Physics, Le Moyne College\\
  Syracuse, NY, USA \\
  \llap{$^e$}Department of Physics, Temple University\\
  Philadelphia, PA, USA \\
  \llap{$^f$}Physics Division, Lawrence Livermore National Laboratory\\
  Livermore, CA, USA\\
  \llap{$^g$}Department of Physics, Illinois Institute of Technology\\
  Chicago, IL, USA\\
  \llap{$^h$}Physical Sciences Laboratory, University of Wisconsin, Madison\\
  Madison, WI, USA\\
  \llap{$^i$}Physics Division, Oak Ridge National Laboratory\\
  Oak Ridge, TN, USA \\
  \llap{$^j$}Department of Physics and Astronomy, University of Tennessee\\
  Knoxville, TN, USA\\
  \llap{$^k$}Department of Physics, Drexel University\\
  Philadelphia, PA, USA \\
  \llap{$^l$}Department of Physics, College of William and Mary\\
  Williamsburg, VA, USA \\ 
  \llap{$^m$}Physics Department, Brookhaven National Laboratory\\
  Upton, NY, USA\\
  \llap{$^n$}Physics Division, Lawrence Berkeley National Laboratory\\
  Berkeley, CA, USA\\
  \llap{$^o$}Chemistry Department, Brookhaven National Laboratory\\
  Upton, NY, USA\\
  \llap{$^p$}National Institute of Standards and Technology\\
  Gaithersburg, MD, USA\\
  \llap{$^q$}Institute for Quantum Computing and Department of Physics, University of Waterloo \\
  Waterloo, ON, Canada \\
  E-mail: \email{ke.han@yale.edu, thomas.langford@yale.edu, blittlej@iit.edu, danielle.norcini@yale.edu}

}
  
\abstract{
A meter-long, 23-liter EJ-309 liquid scintillator detector has been constructed to study the light collection and pulse-shape discrimination performance of elongated scintillator cells for the PROSPECT reactor antineutrino experiment.
The magnitude and uniformity of light collection and neutron/gamma discrimination power in the energy range of antineutrino inverse beta decay products have been studied using gamma and spontaneous fission calibration sources deployed along the cell long axis.  
We also study neutron-gamma discrimination and light collection abilities for differing PMT and reflector configurations. 
Key design features for optimizing MeV-scale response and background rejection capabilities are identified.
}

\begin{document}


\section{Introduction}
\label{sec:intro}
\graphicspath{ {./Figures/Introduction/}}

Since the first observation of the electron antineutrino by Reines and Cowan \cite{Cowan:1992xc}, liquid scintillator detectors utilizing the Inverse Beta Decay (IBD) reaction of $\nuebar + p \rightarrow e^+ + n$  have been the instrument of choice for many reactor \nuebar experiments.  
The delayed coincidence signature of positron annihilation followed by the subsequent neutron capture provides a distinct signature for the identification of \nuebar events and the rejection of accidental backgrounds.

Large-volume, homogeneous liquid scintillator detectors located underground at depths up to 2,700 meters water equivalent (mwe) have been used by the KamLAND, Daya Bay, RENO, and Double Chooz experiments \cite{abe2008KamLAND,An2012DayaBay,Ahn2012RENO,Abe2014DoubleChooz} to make high-precision measurements of reactor antineutrino oscillations and to measure the associated oscillation parameters.  
At distances ranging from one to several hundred kilometers from nuclear power stations, these detectors require active target masses ranging from tens of tons to kilotons and a deep underground location to suppress cosmogenic backgrounds.   

While these reactor \nuebar experiments have successfully measured neutrino oscillations, they have also revealed anomalies with regard to the \nuebar flux and spectrum  emitted by reactors.  
The observed flux appears to be 6\% low when compared to modern predictions of reactor antineutrino production \cite{Mueller:2011nm,Huber:2011wv} and the spectrum reveals deviations from the expected shape in the energy range from 4-6~MeV \cite{Abe2014DoubleChooz, Seo2014RENO, Zhong2014DYB}.  
These observations may point to an incomplete understanding of \nuebar production in nuclear reactors, or to new physics such as meter-scale neutrino oscillation due to the existence of eV-scale sterile neutrinos~\cite{Mention:2011rk, Abazajian:2012ys, Zhang:2013nu}.   
Therefore, new reactor neutrino experiments at baselines of $\lesssim$10~m are being planned to search for indications of short-baseline reactor antineutrino oscillations and make a precise measurement  of reactor-produced $\overline{\nu}_e$ flux and spectra.  
Measurements made by these experiments can provide valuable new insights on the disagreements between the predicted and recently measured reactor spectra.  
In particular, new high-precision measurements of the $^{235}$U \nuebar spectra produced by research reactors with Highly Enriched Uranium (HEU) cores would provide a complementary dataset compared to the existing measurements at power reactors allowing for improved constraints on current and future reactor models~\cite{Dwyer:2014, Hayes:2015}.

Such short-baseline measurements require ton-scale \nuebar detectors with good energy resolution and sufficient position reconstruction to allow observation of sub-meter oscillation wavelengths.  
In contrast to the long-baseline reactor experiments described above, the required reactor-detector proximity in short-baseline measurements almost always requires surface or near-surface operation with minimal overburden.  
For this reason, short-baseline detectors currently under development utilize a variety of design features and detection methods to reduce backgrounds due to the near-surface cosmic ray flux and the reactor environment. 
These include the use of overburden from the reactor building~\cite{DANSS, STEREO}, passive shielding~\cite{PROSPECTwhitepaper,STEREO,Serebrov:2015}, varying levels of detector segmentation~\cite{PROSPECTwhitepaper, DANSS, NuLAT, Solid:2014}, and use of metal-doped~\cite{PROSPECTwhitepaper, STEREO,Serebrov:2015, NuLAT, Solid:2014,Oguri201433} and/or pulse-shape-discriminating (PSD)~\cite{PROSPECTwhitepaper,STEREO,NuLAT,Solid:2014} scintillator.  

$^6$Li-loaded liquid scintillators with PSD capabilities provide a promising avenue for achieving the detector performance necessary for a precise measurement of the reactor antineutrino spectrum and a sensitive search for sterile neutrino oscillations. PSD scintillators have a long history of use in the neutron detection community~\cite{birks}, where cells filled with PSD-capable scintillator are used to identify the high $dE/dx$ proton recoil signatures of incident fast neutrons, which produce a higher proportion of late scintillation light with respect to lower $dE/dx$ electromagnetic energy depositions.
Neutron detectors have traditionally used commercially available liquid PSD-capable scintillators, such as NE-213\footnote{\nistAck}~\cite{Tilquin:1995,Zecher:1997,Ito:1995}, BC501A~\cite{Banerjee:2009}, EJ-301~\cite{Stevanato:2012,Zhang:2013}, or EJ-309~\cite{Kaplan:2013,Cester:2013}, and cell sizes appropriate to the specific application.
For other applications such as time-of-flight spectroscopy or low-background neutron counting, larger cells have been used to increase statistics~\cite{Zhang:2013}, detector surface area coverage~\cite{Zecher:1997}, or neutron detection efficiency~\cite{Tilquin:1995,Ito:1995} at the expense of some neutron-gamma discrimination power~\cite{Banerjee:2009,Moszynski:1994}.
For neutron detection in the presence of high gamma backgrounds, cells with liter-size volumes have been utilized to enhance differences in electromagnetic and hadronic pulse shapes~\cite{Stevanato:2012,Cester:2013}.
Doping of PSD scintillators with $^6$Li allows identification of neutron capture through the heavy ion products of the $^6$Li$+ n \rightarrow \alpha + t$ reaction.
Recent development of lithium-loaded scintillators based on low-toxicity, high-flashpoint organic scintillators with pulse shape discrimination capabilities~\cite{Fisher2011126, Bass2013130, kim2015development, Zaitseva2013747} is particularly important for the operation of liquid scintillator detectors in critical environments such as laboratories near or inside nuclear reactor facilities. 

For the detection of reactor \nuebar at short baselines, the combination of PSD and $^6$Li-doping provides significant discrimination power between signal and background.  
Antineutrino-produced IBD \betaplus{}-neutron pairs produce sequential gamma-like and neutron capture-like energy depositions.
This pattern is clearly distinguishable from the most copious near-surface IBD backgrounds from cosmogenic fast neutrons, which produce sequential neutron-like signals from proton recoil and subsequent n-$^6$Li capture, and from accidental coincidence of sequential ambient gamma rays.
This scintillator deployed in a close-packed array of optically isolated cells can provide the large segmented target volume needed for good IBD event statistics, efficiency, position resolution, and background rejection, as demonstrated by the Bugey~3 experiment~\cite{Abbes:1996}, which used an obsolete lithium-loaded pseudocumene-based liquid scintillator~\cite{Ait:1998}.

Achieving the superior energy resolution required for constraining current reactor \nuebar production models will require advances beyond currently demonstrated segmented detectors, including a low inactive volume, high photon yields and collection efficiencies, and excellent energy response uniformity.  
A number of new technological developments can be utilized to meet these requirements. Some recently-developed $^6$Li-doped PSD scintillators have demonstrated excellent optical photon yields~\cite{norcini2015LiLS}.
Highly efficient diffuse and specular reflectors developed for use in the lighting and solar collection industries have the potential to greatly improve photon collection efficiency.
Finally, segmentation systems utilizing a single thin wall between adjacent segments can greatly reduce the inactive detector mass fraction from the $\sim$15\% level achieved by Bugey~3 down to the percent level.  

The PROSPECT experiment, an effort to measure the $^{235}$U \nuebar spectrum from the High Flux Isotope Reactor (HFIR) at the Oak Ridge National Laboratory over a range of short baselines, has been proposed and designed with these detector requirements and technological developments in mind.
PROSPECT will utilize a rectangular $^6$Li-loaded liquid organic PSD scintillator target optically segmented into 1.2~meter-long rectangular cells with read-out on two ends by PMTs~\cite{PROSPECTwhitepaper}.
The optical segmentation system will consist of an array of  millimeter thick, double-sided reflecting panels held in position by reflective support rods.
The segmentation system will mate directly to an optically transparent mirrored light guide coupled to each PMT, providing a continuous reflective surface within each scintillator cell.
Time-correlated IBD \betaplus and n-$^6$Li energy depositions produce scintillation light, which will be guided with high efficiency to the PMTs via reflections off the separator system walls and rods.  
Pulse shape discrimination of PMT waveforms will allow separation of neutron captures from background electromagnetic interactions and prompt positron signals from proton recoils due to cosmogenic fast neutrons.  
Double-ended PMT readout will provide the further ability to infer event positions along the cell length.

This paper describes tests performed with a 23-liter, 1-meter long test detector using EJ-309 scintillator. This detector is used as testbed for the study of detector elements used in the PROSPECT detector. We examine the total magnitude of achievable light collection for varying configurations using gamma and spontaneous fission sources.
We also measure PSD performance for these varied configurations as well as examining the correlated and uncorrelated aspects of the PSD and light collection.
Variations in performance metrics along the length of the test detector are also investigated, with the goal of identifying a configuration with minimal variation in these parameters over the meter-long detector length.  
The results will guide the final design of the segmented PROSPECT detector and large liquid scintillator detectors for other applications.

Section~\ref{sec:detector} describes the design and construction of the scintillator cell, its main components, and achievable variations of the deployed cell geometry.  
Acquired datasets from the cell and subsequent analysis methods and metrics are described in Section~\ref{sec:Data}.
Section~\ref{sec:Results} presents results describing the light collection and PSD performance metrics examined in this study using a `default' configuration described in that section.
Section~\ref{sec:designOpts} describes datasets taken with alternate test detector reflector and PMT configurations and attendant observed variations in detector response.
Finally, Section~\ref{sec:Interior} describes cell alterations made based on these studies to produce a more PROSPECT-like geometric and optical design and describes the improved performance of this configuration.
Implications and key findings are summarized in Section~\ref{sec:conc}.

\section{Cell Design and Construction}
\label{sec:detector}
\begin{figure}[h]
\begin{center}
\includegraphics[width=0.75\textwidth]{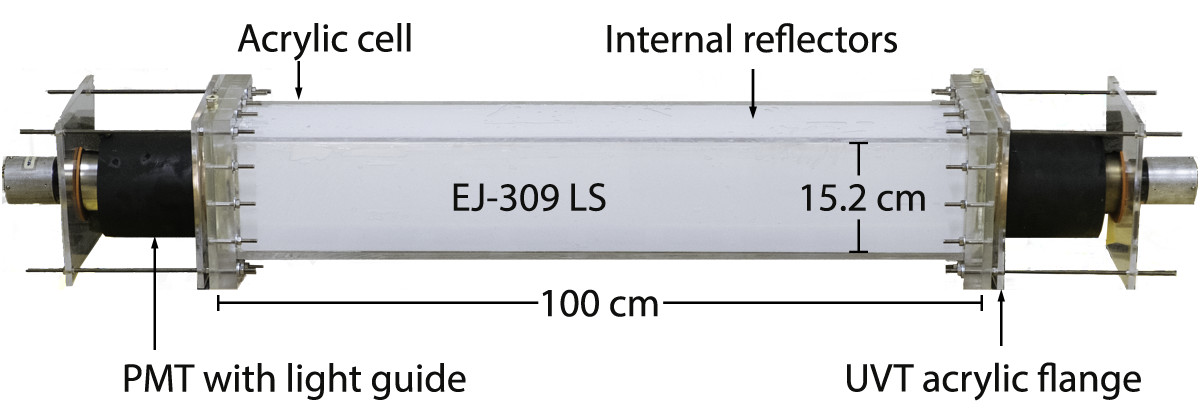}\\
\includegraphics[width=0.9\textwidth]{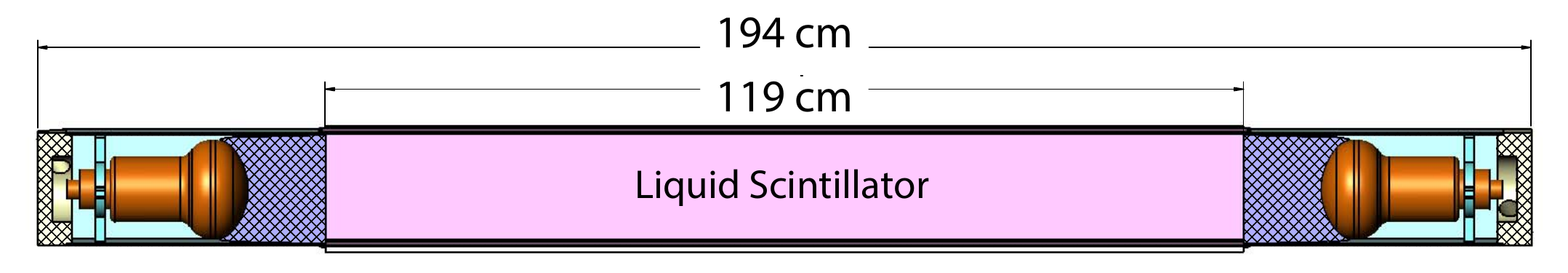}
\caption{Top: A labelled photograph of the test detector.  The picture shows a test detector configuration featuring internal reflectors, which are described in further detail in Section~6.
Bottom: A schematic of the baseline PROSPECT design for a single optical cell.
}

\label{fig:detectorSchematic}
\end{center}
\end{figure}

\graphicspath{ {./Figures/detector/}}
The test detector, shown in Figure~\ref{fig:detectorSchematic}, has primarily been produced to study the design parameters and response of the PROSPECT detector's cell geometry. 
The test detector's design reflects many of the essential features of the PROSPECT cell design:
\begin{enumerate}
\item{A long cell aspect ratio.}
\item{Reflecting walls along the cell length.}
\item{Matched PMT and cell cross-sections.}
\item{Optical coupling of cell-end PMTs.}
\item{Commercial PMT, cable, high voltage (HV), and data-acquisition (DAQ) models.}
\end{enumerate}

The test detector described here differs from current baseline PROSPECT cell design in a variety of aspects.  Items listed below give the differing design feature of PROSPECT versus the test detector, in that order.
\begin{enumerate}
\item{Cell length: 1.2 versus 1.0 meters.}
\item{Cell wall composition: mm-level thickness separators versus cm-thick acrylic cell walls.}
\item{Wall-PMT mating: direct mating versus gaps in reflective surfaces due to cell flanges.}
\item{PMT light guide shapes: parabolic versus cylindrical.}
\item{Scintillator choice: custom Li-doped scintillator versus un-doped EJ-309.}
\end{enumerate}

The test detector has also been designed to allow for variation of key cell components to inform PROSPECT optical cell design choices: the studies described below include variation in PMT deployments and in reflector types, locations, and coupling methods.
The following sections describe in further detail the test detector's main components and implemented variations.  

\subsection{Acrylic Cell}

The main structural component of the test detector is the acrylic cell, which consists of a long rectangular 1~cm-thick UV-absorbing (UVA) acrylic barrel with a 2.5~cm thick square flange adhered to either end.
UVA end plates of 2.5~cm thickness are bolted to the flanges with a Viton o-ring to seal the cell.
The flanges provide access to the cell interior, which can be cleaned via an Alconox solution and rinsed before filling.
This design also allows the insertion of internal reflectors, described in Section~\ref{sec:Interior}.
Acetel valves with Viton o-rings are located on each end plate corner for filling and gas purging.
With no interior components installed, the active volume is 15.2 x 15.2 x 100~cm$^3$.
The only materials in direct contact with scintillator in this configuration are acrylic and small amounts of Viton and Acetel.

\subsection{Liquid Scintillator}

For this work, the test detector's acrylic cell was filled with EJ-309, a non-toxic, non-flammable, liquid scintillator from Eljen Technology designed for use in pulse shape discrimination applications~\cite{EJ309}. 
The manufacturer quotes a light yield of 11,500~photons/MeV and an emission spectrum dominated by wavelengths above 375-400~nm, the characteristic transmittance cut-off range for UVA acrylics.
Previous work has characterized the light collection and neutron-gamma discrimination abilities of EJ-309~\cite{Kaplan:2013,Cester:2013}.
However, most of these studies have been limited to smaller (sub-liter or few-liter) volumes.
While custom Li-doped EJ-309 scintillator has been developed by the PROSPECT collaboration~\cite{norcini2015LiLS}, un-doped EJ-309 was utilized in this study because its properties are well-documented in the literature.

The properties of the scintillator were purposefully not altered during all tests described in this paper.
Special care was taken during cell filling to reduce exposure to the ambient laboratory atmosphere, so as to maintain the high performance and stability.  
The scintillator was manipulated during filling inside a glove box filled with high-purity nitrogen gas.  
Scintillator was delivered to the pre-purged acrylic cell via a peristaltic pump and teflon tubing.  
After filling, the cell filling port was temporarily sealed, and the cell was maintained at constant temperature with variations of less than 3$^{\circ}$C.

\subsection{Reflectors}
A significant fraction of visible photons produced by scintillation are transported to the end-mounted PMTs via reflections along the length of the cell.
This propagation is supported by either total internal reflection (TIR) at detector material boundaries or reflection from highly reflecting surfaces along the cell length.

To examine these contributions independently, different optical surfaces were applied to the outside of the acrylic cell.
Exterior application has the advantage of allowing rapid testing of reflector types without disturbing other cell components.  
However, this comes at the expense of non-optimal matching of the PMT photo-sensitive diameter (11~cm) and the enclosed optical cell dimension (17.2~cm).
Poor reflectors composed of black cloth were wrapped along the exterior length of the acrylic cell to determine detector properties with only TIR produced at the acrylic (n$\approx$1.48) - air (n$\approx$1.0) interface on the cell's outer surface.
Highly specular reflectors made from either 3M SolarMirror 2020~\cite{SM2020} or 3M Enhanced Specular Reflector~\cite{ESR} film were attached to the cell exterior without any index-matching substance (`un-coupled') to determine response with TIR combined with collection of light at high incident angles.
Similar tests were also done with WhiteOptics White98 diffuse reflectors~\cite{White98}.
In some tests, the specular and diffuse reflectors were optically coupled to the cell exterior utilizing glycerin (n$\approx$1.47), greatly reducing TIR and allowing investigation of light collection due to the reflective material.
To visually verify the completeness and reproducibility of the couplings, only two of the four deployed reflectors were glycerine-coupled at any given time.
In all cases, a non-trivial amount of TIR also occurs at the scintillator~(n$\approx$1.56) - acrylic (n$\approx$1.48) interface.  
It is also noted that the cell end flanges are not covered in reflective material, leaving two 5~cm gaps in the length of the cell through which scintillation light can escape.  

Based on the investigation of these differing reflective cases with externally-applied reflectors, internal reflectors that sit inside the acrylic cell in contact with the EJ-309 scintillator were developed and deployed.  The composition, design advantages, and performance of these internal reflectors are described in Section~\ref{sec:Interior}.

\subsection{PMTs and Endcaps}
Light is detected via Hamamatsu R6594 12.7~cm diameter hemispherical-faced PMTs (11~cm photocathode diameter) coupled to the acrylic endcaps of the cell.
Manufacturer specifications indicate a difference in quantum efficiency of less than 1\% between the two PMTs used.
Custom UVT acrylic optical couplers of cylindrical shape with 11.4~cm radius and 5~mm minimum thickness were produced and coupled to the PMTs utilizing RTV-615 silicone. 
Optimization of the coupler shape may yield additional light collection efficiency.  
PMT-coupler assemblies were rigidly mounted to the cell ends with a custom plate secured to the cell via threaded rods.
The PMT-coupler combination was then coupled to the acrylic cell via RTV.
This allowed for the detector to be handled and moved without disturbing the optical coupling of the PMTs.
Both PMTs were surrounded with mu-metal to shield them from magnetic fields.

\subsection{Electronics}
High voltage was provided by CAEN V6533 VME-based power supplies. 
Signals from both PMTs were sent via 10~m cables to a CAEN 1730 waveform digitizer (16~channel, 14~bit dynamic range, 500~MS/s sampling frequency) and analysis was performed offline using the IPython scientific programming suite~\cite{PER-GRA:2007, Hunter:2007}.  
A CAEN V1720 waveform digitizer (8~channel, 12~bit dynamic range, 250~MS/s sampling frequency) was also utilized to investigate the effect of reduced electronics performance on PSD.
The slower digitizer did not result in significantly reduced PSD performance, though the reduced dynamic range limited its functionality.
For this reason, all data reported here was produced using the V1730 digitizer triggered internally by either PMT with a threshold of 50~ADC~channels, or 25~mV.


\section{Data and Analysis}
\label{sec:Data}
\graphicspath{ {./Figures/analysis/}}
A variety of calibration sources were deployed along the cell exterior to study the detector response.  
A \Bi gamma source was utilized primarily for photoelectron (PE) collection characterizations, while a \Cf spontaneous fission source was used for neutron studies.  
During a data-taking run, a source was placed at one position along the long (z) cell axis.
For some runs, as indicated in the text, the source is collimated into a fan-shaped beam oriented perpendicular to the long cell axis by using two lead bricks separated by 5~mm.  
The \Cf source is deployed at a distance of $\sim$50~cm from the side of the cell, while the \Bi source was placed directly on top of the cell or, when utilized, the collimating bricks. 
For  \Cf, a neutron collimator was used to reduce the fraction of `room-return' neutrons, while a lead shield was used to reduce the total number of gamma interactions in the cell.  
In all runs, a 5~cm thick layer of lead was placed around the full detector to reduce overall trigger rates and impact from background radiation.

A CAEN SP5601 low-intensity LED pulser was used to provide a high-purity sample of single-PE calibration data.
By fitting these data, ADC-PE conversions were determined for each PMT at their nominal operating voltages.
The digitized signals are then integrated and divided by each PMT's single-PE value, yielding the number of PE detected per event.
By comparing the number of PE detected for each longitudinal source position, the light-collection uniformity can be studied.

\begin{figure}[htb!]
\begin{center}
\includegraphics[width=.48\textwidth]{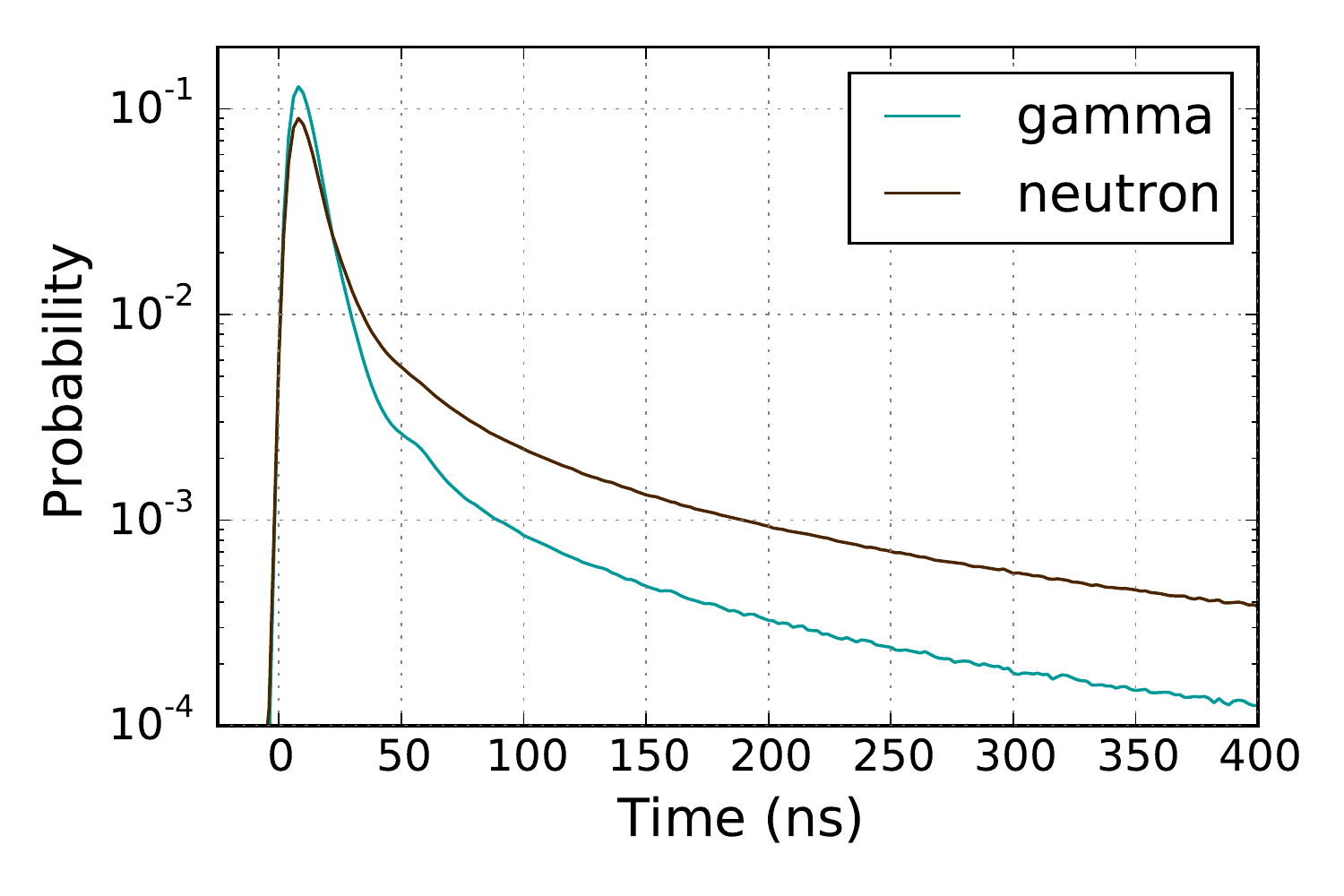}~
\includegraphics[width=.48\textwidth]{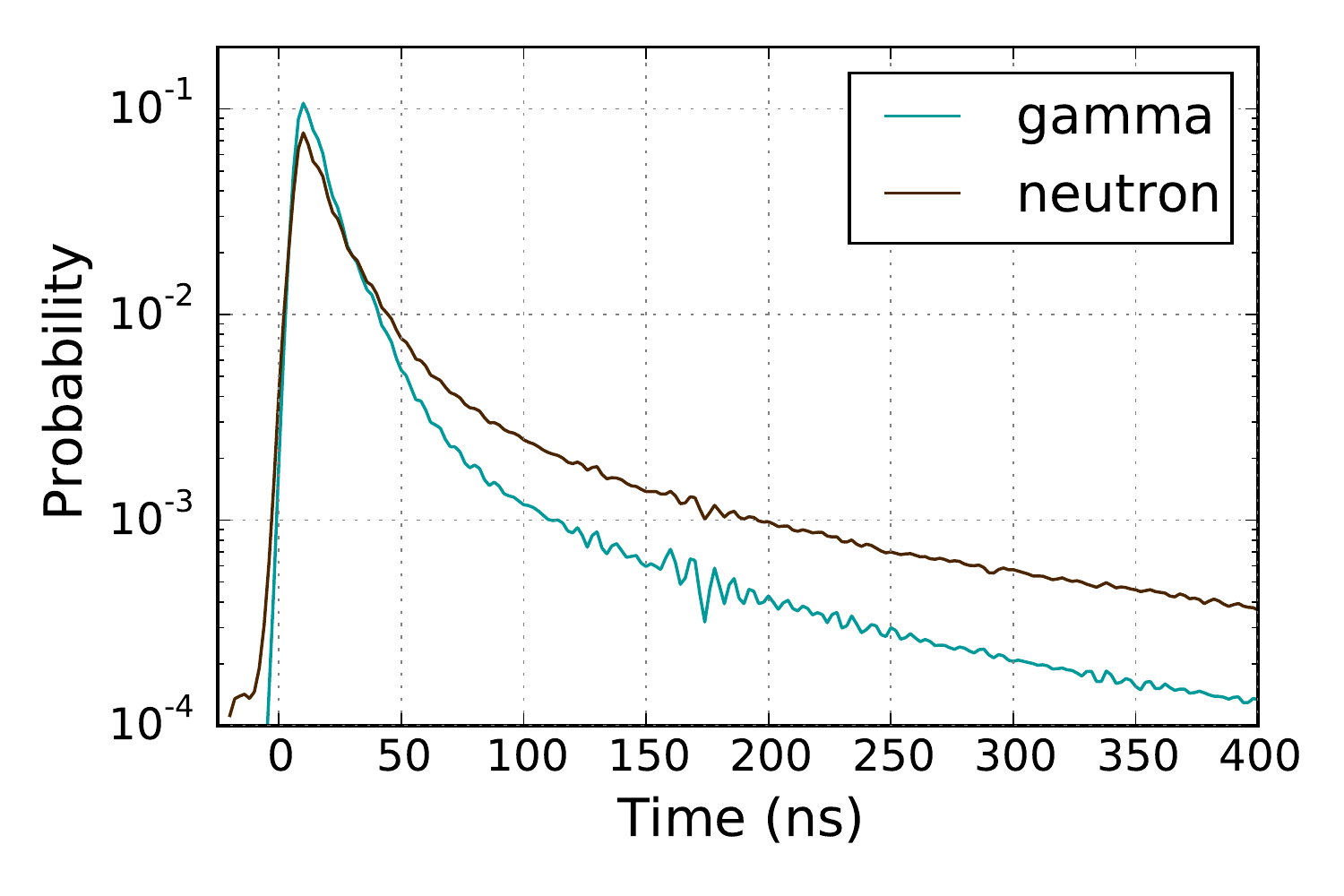}
\caption{Template pulses derived by averaging signals from neutron and gamma interactions in a 100-milliliter cell of EJ-309 (left) and the test detector (right). 
Though the large-volume pulses have been broadened by geometric effects, significant PSD discrimination capability is retained. 
Small features are present in the tails of each template due to slight impedance mismatches between cables.
}
\label{fig:PSDSchematic}
\end{center}
\end{figure}

The time structure of each recorded PMT waveform is analyzed using the `tail-fraction' pulse shape discrimination metric.  
Since neutron and gamma interactions have different characteristic decay-times, the fraction of light collected in the tail of a signal can be used to isolate each event type~\cite{Pawelczak:2013}.  
Figure~\ref{fig:PSDSchematic} shows pulses generated by averaging signals from neutron and gamma interactions in a 100-milliliter cell of EJ-309 and in the test detector. 
Though the behavior of the slow component is similar between the two setups, there is a significant broadening in the fast component of the signal, presumably due to dispersion resulting from the larger range of propagation lengths in the larger geometry. 
This difference highlights the need to quantify PSD performance in elongated scintillator cells, and to optimize PSD metric calculations specifically for differing cell geometries.

As a measure of PSD performance, the figure-of-merit (FOM) is determined by fitting the PSD distribution with two Gaussians and extracting the means ($\mu$) and full width half maxima (FWHM):
\begin{equation}
FOM = \frac{\mu_2 - \mu_1}{FWHM_1 + FWHM_2}.
\end{equation}
For each experimental setup, optimization of the PSD metric is performed by incrementing through a series of integration ranges.
The integration region that provides the best FOM is selected for use.
The moderately non-Gaussian behavior of PSD distribution tails produces some instability in fit widths, resulting in a 5\% fitting uncertainty in the PSD FOM.
We also note that the re-optimization of PSD FOM produces differing PSD peak values for configurations of widely differing PE yields.


\section{Performance Demonstration with the Default Configuration}
\label{sec:Results}
\graphicspath{ {./Figures/results/}}

We now consider what we have designated the `default' test detector configuration.
All other tested configurations described in the paper differ from this default configuration in only one well-defined design aspect.
This configuration consists of the described acrylic cell and EJ-309 scintillator accompanied by:

\begin{itemize}
  \item Double-ended PMT readout with coupled Hamamatsu R6594 PMTs.
  \item Un-coupled 3M ESR specular reflectors covering the entire exterior of all four cell sides.
  \item Un-coupled 3M ESR specular reflectors covering the sides of the cylindrical PMT optical coupler.
  \item CAEN 1730 waveform digitizer.
\end{itemize}

\subsection{Absolute Light Collection}

\begin{figure}[h!]
\begin{center}
\includegraphics[width=0.48\textwidth]{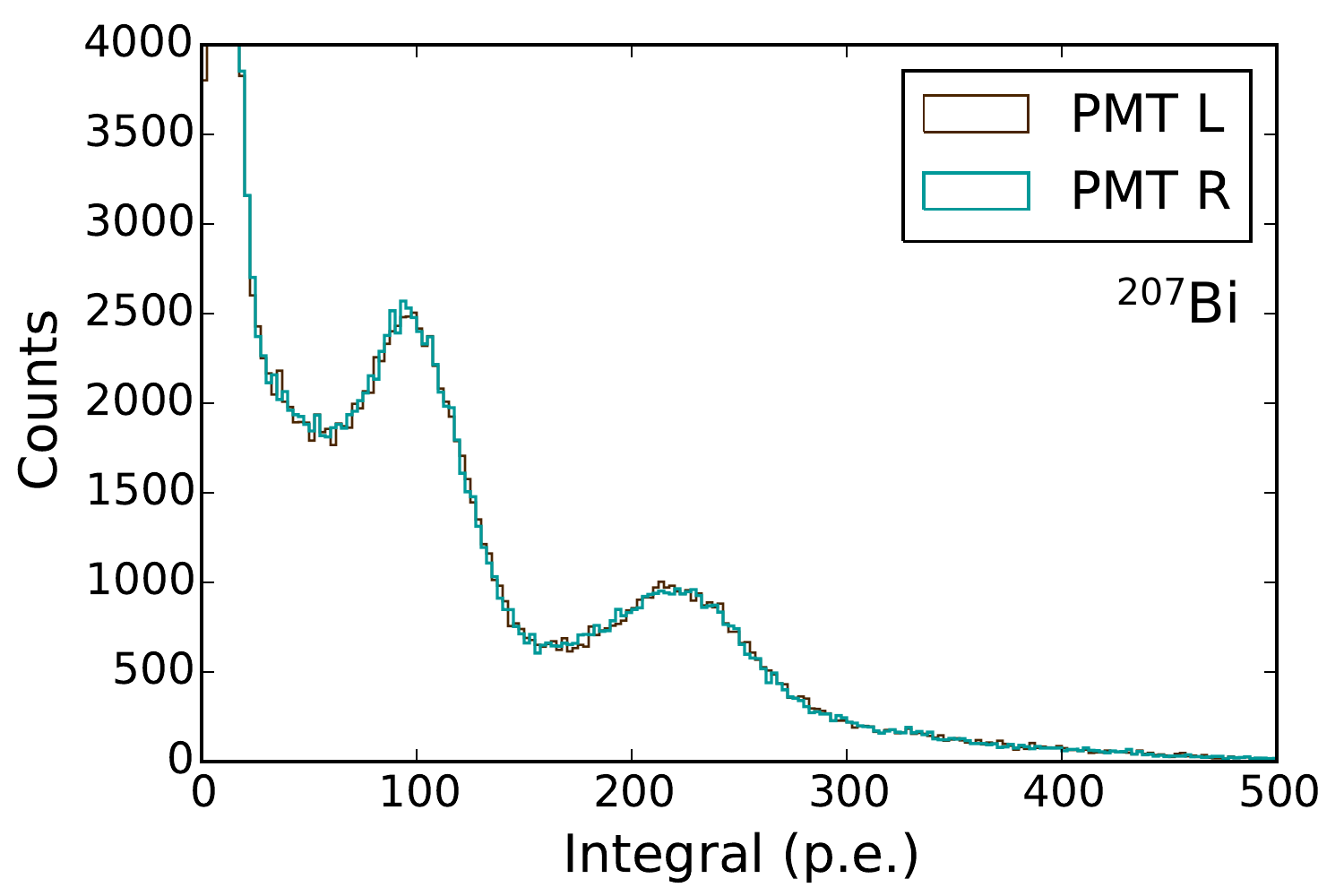}~
\includegraphics[width=0.48\textwidth]{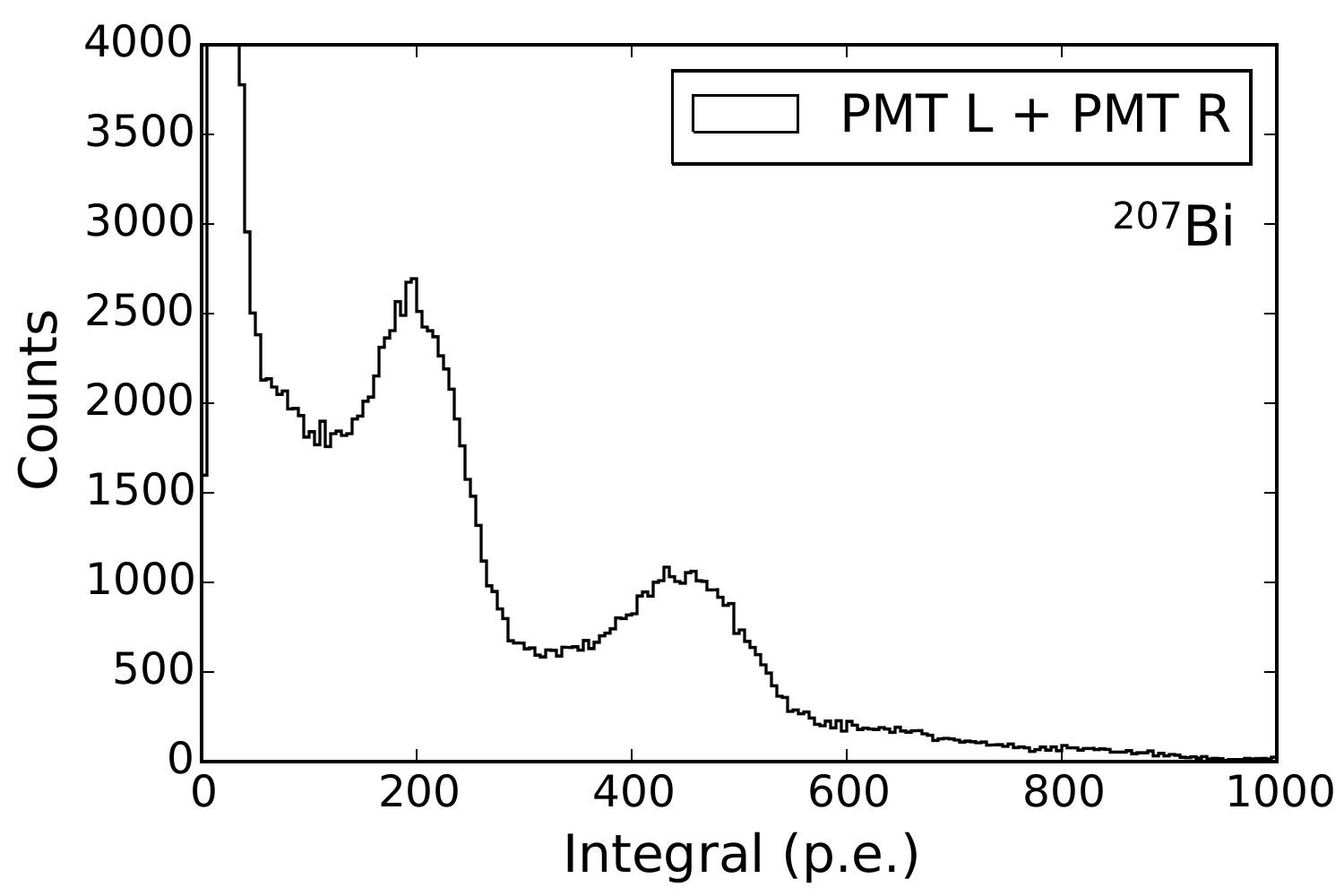}
\includegraphics[width=0.48\textwidth]{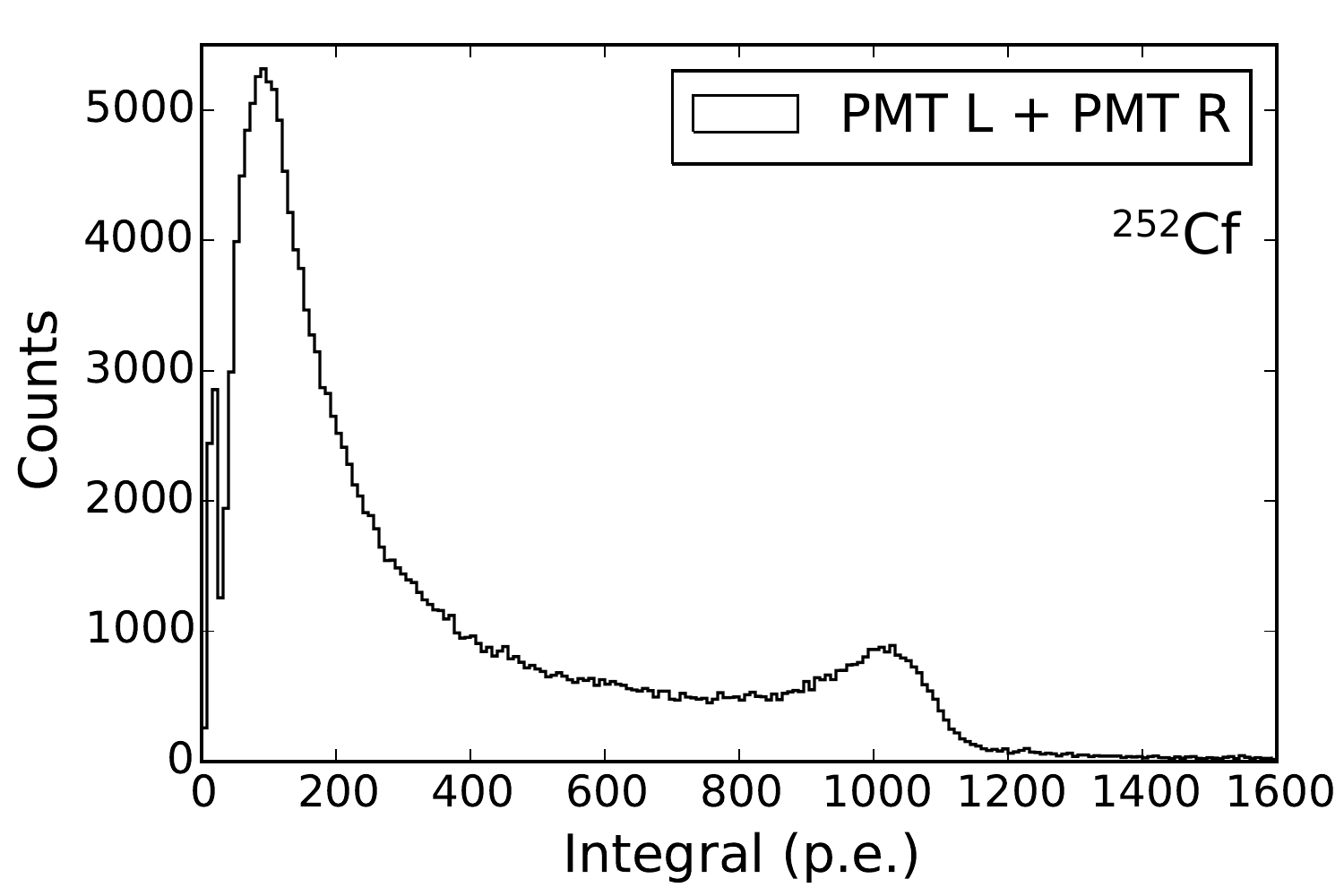}~
\includegraphics[width=0.48\textwidth]{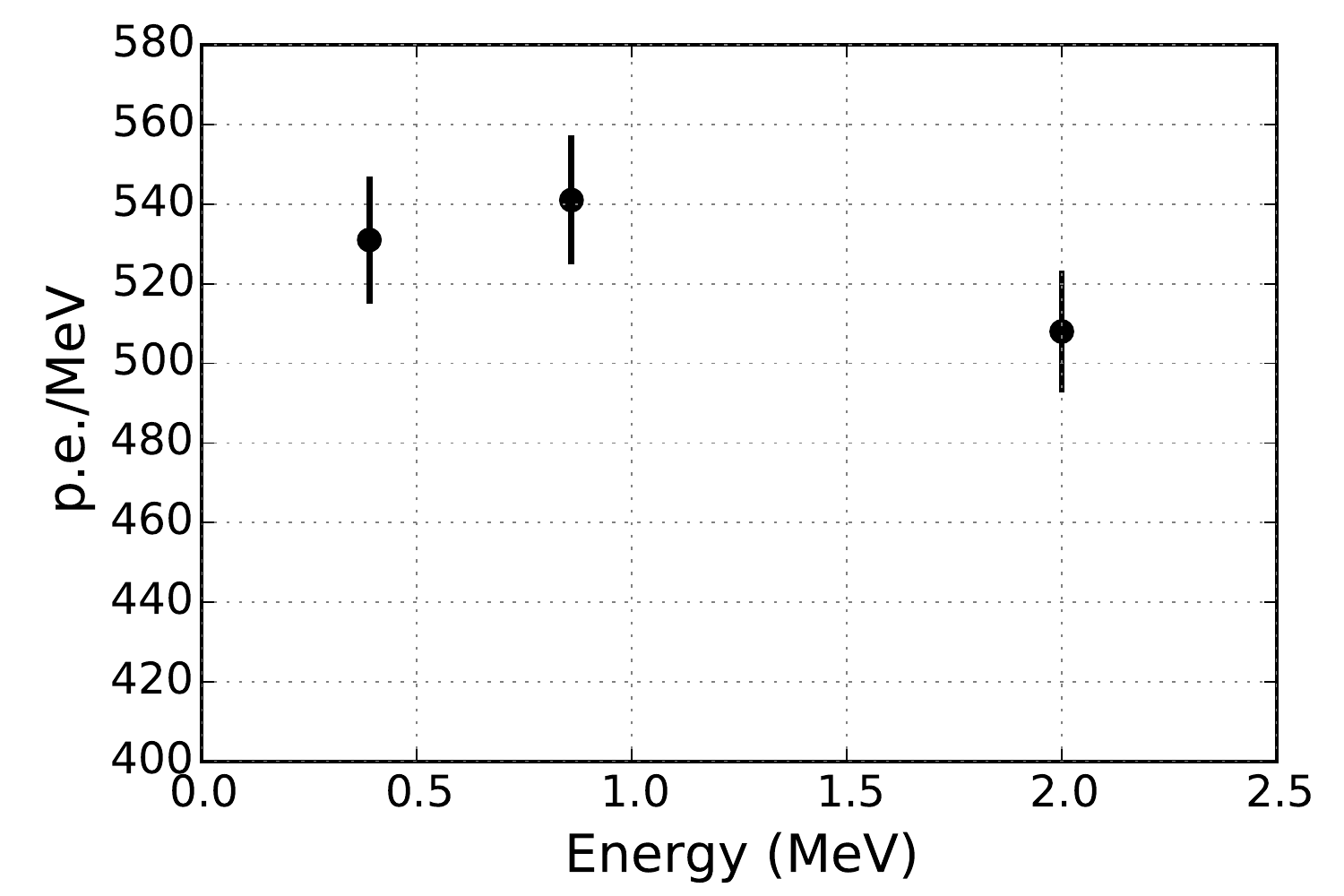}
\caption{
Top: Calibrated PE spectra of left and right Hamamatsu PMTs (left) and summed PE (right plot) in the default cell configuration in response to center-deployed \Bi.
Bottom Left: Summed PE spectrum from a center-deployed \Cf source. 
Bottom Right: Distribution of the PE/MeV values obtained by fitting the three gamma lines. 
Reasonable agreement between these values is visible within the fitting uncertainties described in the text.
}
\label{fig:hamamatsuPE}
\end{center}
\end{figure}

Figure~\ref{fig:hamamatsuPE} shows the PE spectra from the left (L) and right (R) Hamamatsu PMTs for the default configuration produced by a \Bi external source deployment at the cell midpoint ($z=0$~cm).
One can see peak-like features in the integral spectrum arising from electron capture gamma-ray lines at 0.57~and~1.07~MeV, with roughly equal light collection given by L and R PMTs.
PE spectra obtained by combining the light detected in both L and R PMTs, referred to as L+R, are also shown in Figure~\ref{fig:hamamatsuPE} for \Bi as well as \Cf source deployments.
The peak at roughly 1000~PE in the \Cf spectrum is from neutron capture on hydrogen, which produces a 2.2~MeV gamma ray.

The dimensions of the cell are comparable to 1~MeV gamma interaction lengths, so a mixture of single and multiple Compton scatters contribute to the peak-like features observed for each gamma line from a calibration source. 
Therefore, neither a Compton edge nor the full energy can be used to calibrate the energy scale.
The total PE yield of the detector was determined by comparing the detected energy spectra to smoothed deposited-energy spectra produced using a simple Geant4 simulation of the scintillator cell.
In both data and simulation, the features were fitted with a Gaussian plus exponential curve to determine the central value, which then define a PE/MeV value for each gamma-ray line. 
Having identified the corresponding mean energy of the features caused by mono-energetic gamma ray lines, we will hereafter refer to such features as ``peaks'' for simplicity.

The described fits produce L+R PE peak values of $202.0\pm0.2$~PE, $459.9\pm0.4$~PE, and $1016.8\pm0.5$~PE for the \Bi 0.52~MeV (0.38~MeV), \Bi 1.07~MeV (0.85~MeV), and \Cf 2.2~MeV (2.04~MeV) gamma lines (peak values), corresponding to light collection of 531, 541, and 508~PE/MeV, respectively.
Dominant uncertainties in PE yield comparisons between source locations arise from instabilities and variations in fit values depending on the applied fitting range, which was of order 2\%.
Statistical uncertainties are of order 0.1\%. 
Non-source triggers from ambient backgrounds provided $<$5\% of the trigger rate, and had a negligible effect on the light collection measurements.

\subsection{Z-Dependence of Light Collection}
\label{sec:lightCollectionZdep}

\begin{figure}[h]
\begin{center}
\includegraphics[width=0.48\textwidth]{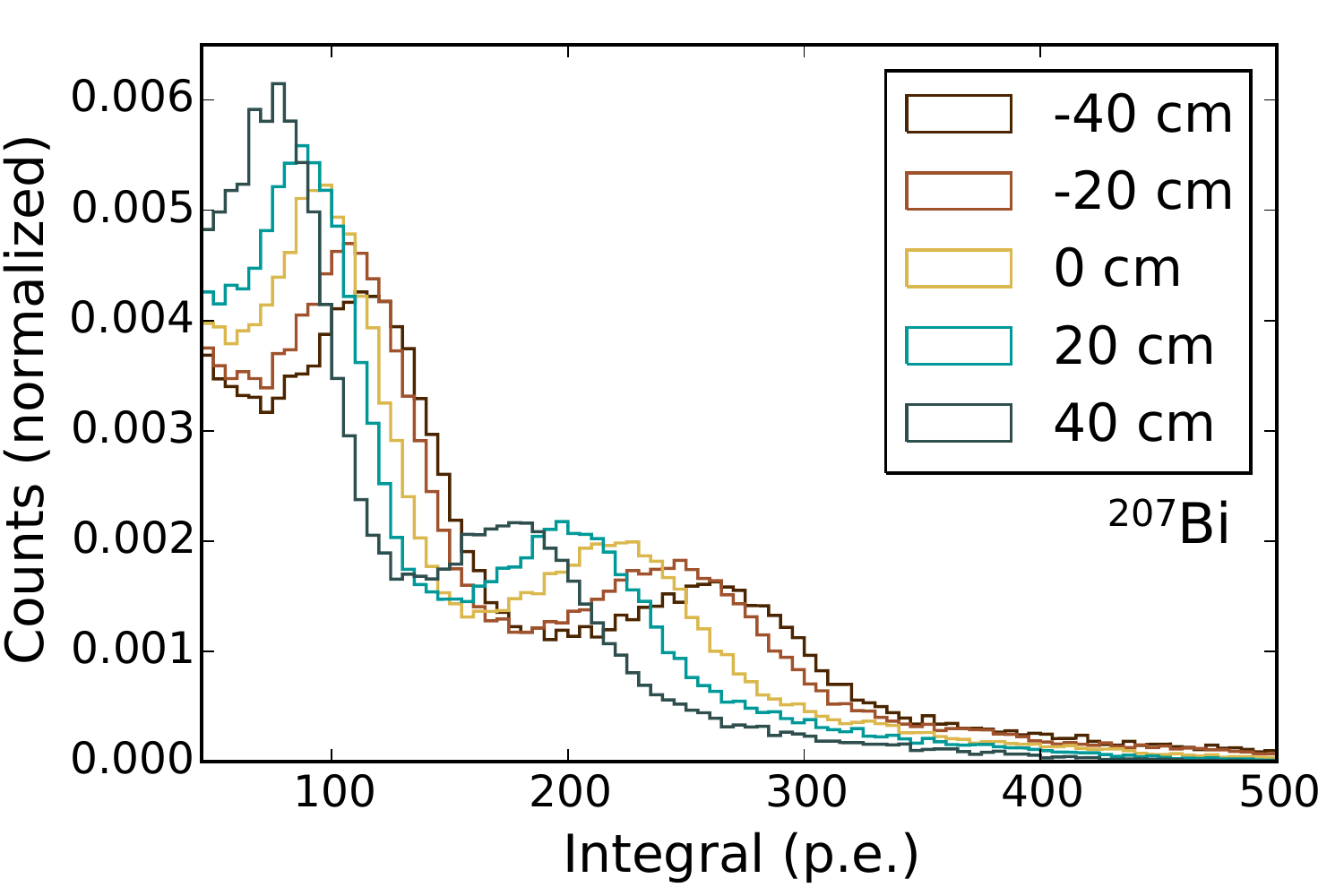}
\includegraphics[width=0.48\textwidth]{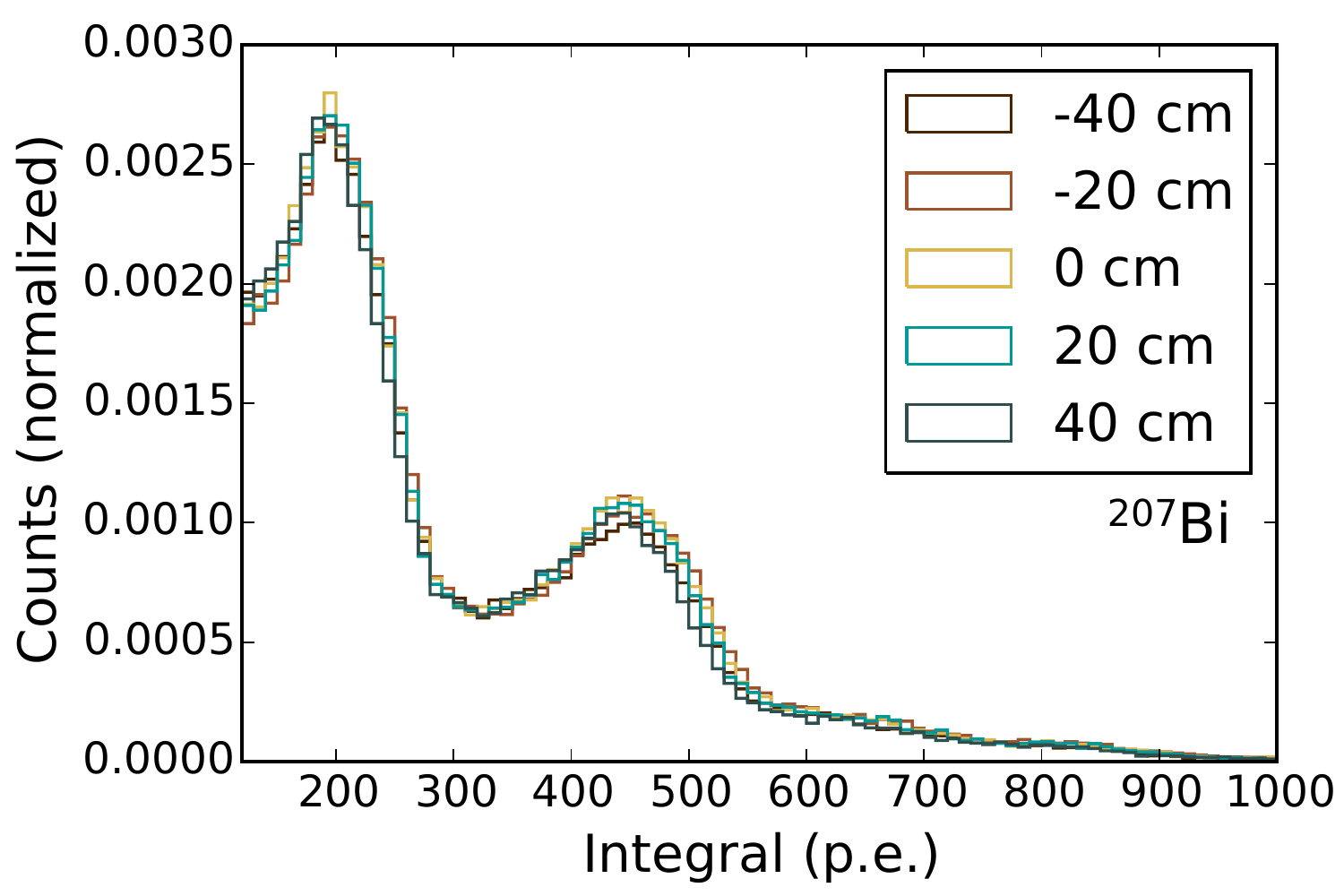}
\caption{PE spectra for PMT L (left) and PMT L+R (right) in response to \Bi source deployments at a range of z-scan locations.  
Excellent uniformity in light collection along the cell is maintained in an elongated EJ-309 cell when information from PMTs at both cell ends are utilized.  
}
\label{fig:baselineLightLR}
\end{center}
\end{figure}

We have also characterized the variation in collection efficiency for particle interactions at various points along the cell length. 
The PE spectra for the R PMT in the default configuration with a \Bi source deployed in 20~cm increments along the cell are shown in the top of Figure~\ref{fig:baselineLightLR}.  
The positions along the cell are oriented such that the L PMT is located closest to -40~cm, with R PMT closest to 40~cm.
For the off-center positions, one can see a clear shift in peak positions for the L PMTs resulting from shorter path-length and higher direct-light solid angle in the closer PMT.
A similar pattern is observed in the R PMT.
For the L+R PE sum, one obtains the charge distributions pictured in the right of Figure~\ref{fig:baselineLightLR} for the same source deployment locations, which exhibit a high degree of uniformity.
Though the scintillator manufacturer only quotes the attenuation length as >1~m, the observed uniformity points to a significantly longer attenuation length. 

Light collection values were determined for each source deployment location and for each PMT separately and combined, using the fit method described above.
For the L+R combination, the peak positions vary by less than 2\% from cell center to end, in contrast to the 40\% variation for individual PMTs.

\subsection{Position Reconstruction}
\label{sec:zReconstruction}
By utilizing either relative timing or charge division between PMTs, two independent estimators of interaction position in the test detector can be produced. 
This ability can be used to increase a detector's rejection power of various classes of backgrounds. 
It also allows the correction of any residual light collection non-uniformities that are a function of longitudinal position.

\begin{figure}[h]
\begin{center}
\includegraphics[width=0.48\textwidth]{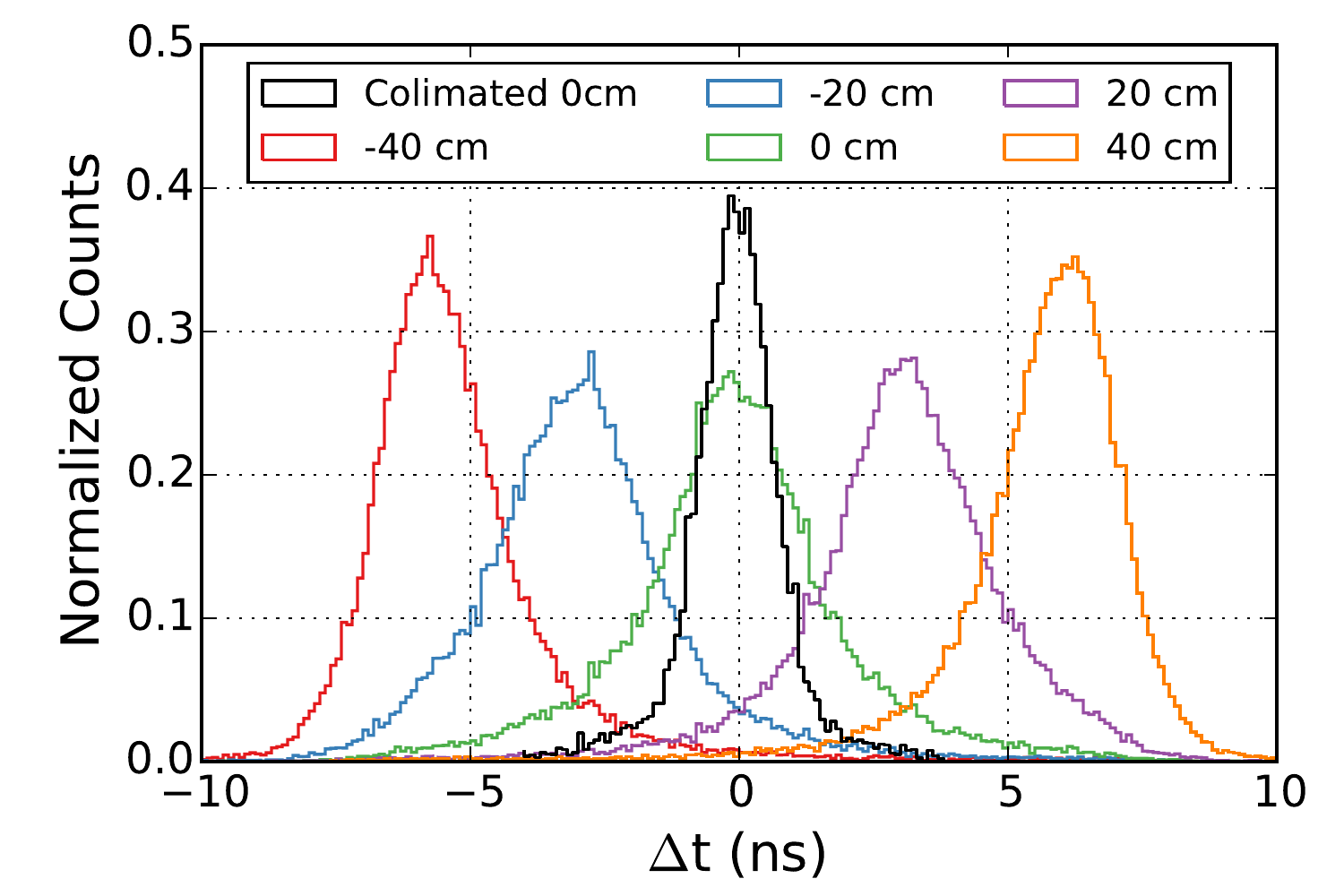}
\includegraphics[width=0.48\textwidth]{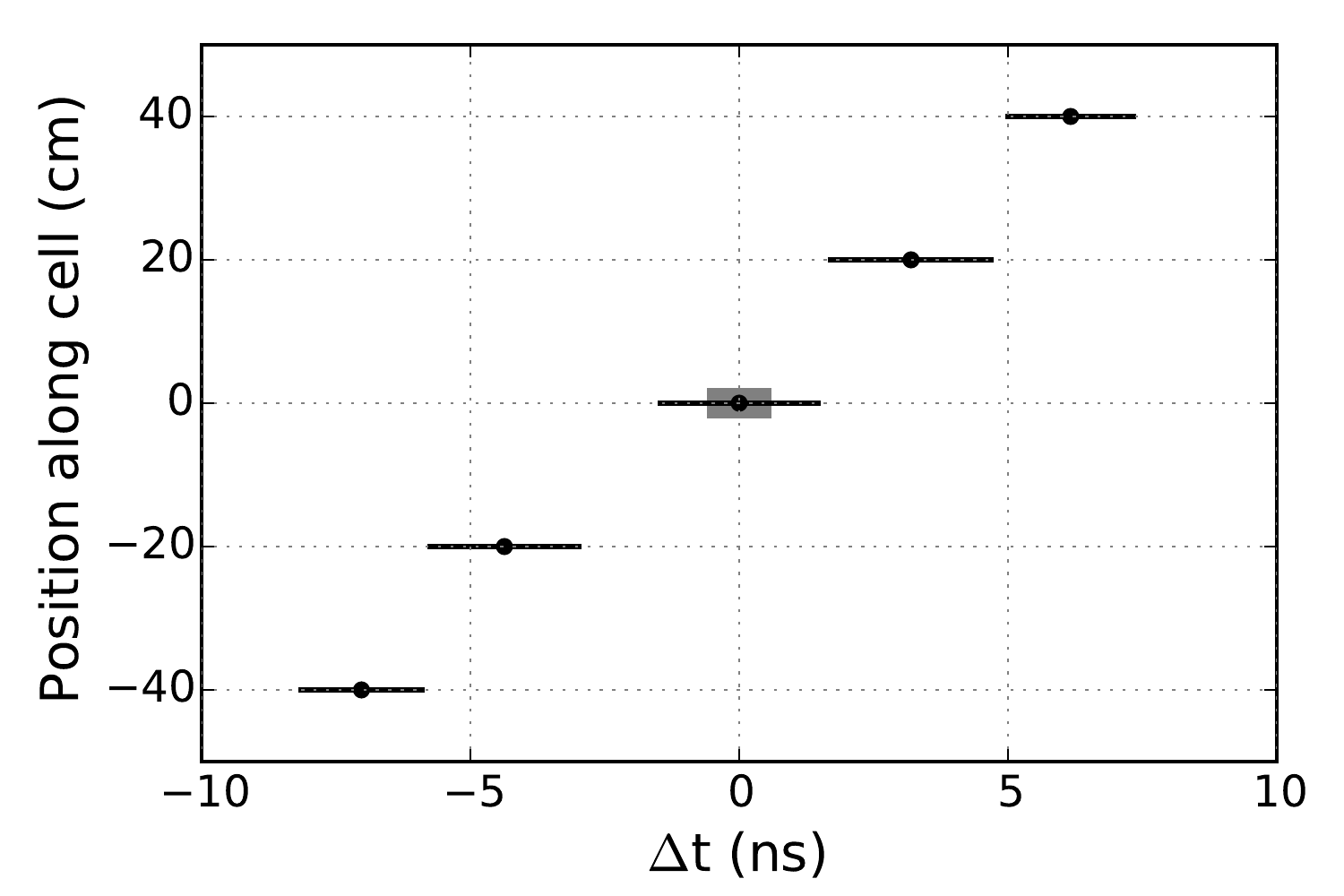}

\caption{Left: The time difference (in nanoseconds) between the two PMT signals for all triggers between 0.5 and 1.0 MeV at various uncollimated \Bi z-deployment positions, as well as one at 0~cm using a lead collimator.
Right:  The fitted Gaussian average and 1-$\sigma$ widths from each uncollimated source deployment. 
The shaded band on the $z=0$~cm point demonstrates the improved width obtained when collimating the source.
}
\label{fig:recoPosition}
\end{center}
\end{figure}

Due to the extended geometry of the cell and the high sampling frequency of the digitizer, the location of an interaction can be reconstructed using the difference in photon arrival times in the two PMTs.
The arrival time was determined by interpolating the time at which the PMT waveform arrives at its half-height.
Timing offsets due to variations in cable length and PMT transit times are removed using the cell-midpoint ($z=0$~cm) \Bi deployment, which should exhibit identical cell-end arrival times. 
The mean and width of the distribution of arrival times are extracted by fitting a Gaussian to \Bi calibration events with energies between 0.5 and 1~MeV at individual z-deployments.  
To determine the position resolution, a collimated \Bi source was deployed at the cell center.\footnote{The collimated source was difficult to reposition, so only the center position was measured.}
A fitted Gaussian width of $\sigma=0.6$~ns was obtained, equivalent to a resolution of 5~cm.
Figure~\ref{fig:recoPosition} shows the arrival time distributions from each source position, including the collimated source. 
The relationship between PMT timing offsets and deployed z-location for each run are also shown. 
A linear relationship between these two quantities is clearly visible in the data.

In addition to this timing reconstruction, a direct relationship also exists between PMT integrated charge and distance from PMT to particle interaction location.  
For double-ended PMT readout, this relationship can be exploited by constructing the following metric utilizing information from both PMTs~\cite{Classen2015139, Zhang:2013}:
\begin{equation}
Z(q) = \log\frac{q_{PMT1}}{q_{PMT2}}.
\label{eq:logratio}
\end{equation}
This metric depends on the PE statistics of the two PMT signals, and so depends on the total light collection efficiency and the chosen energy range.

For each source-deployment position, the width of the distribution is found to be $\sigma$ = 0.065, equivalent to 16~cm position resolution in the meter-long cell.  
For collimated data, this resolution is improved to 14.5~cm.
The lower resolution for this method is due to the relatively efficient light propagation achieved in the cell, which reduces the contrast between the L and R integrals as a function of position.

\subsection{Pulse Shape Discrimination Performance}

\begin{figure}[htb!]
\begin{center}
\includegraphics[width=0.48\textwidth]{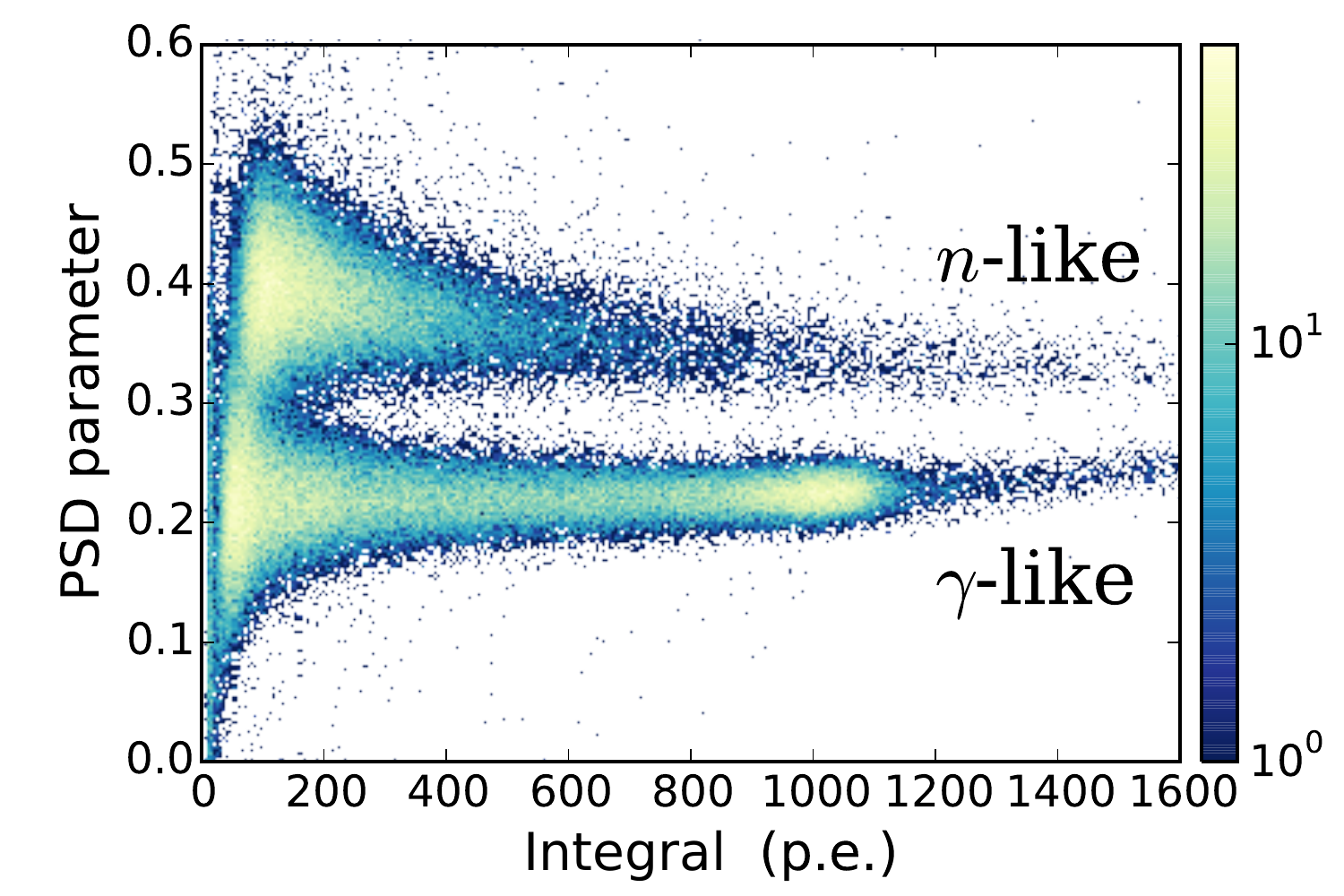}
\includegraphics[width=0.48\textwidth]{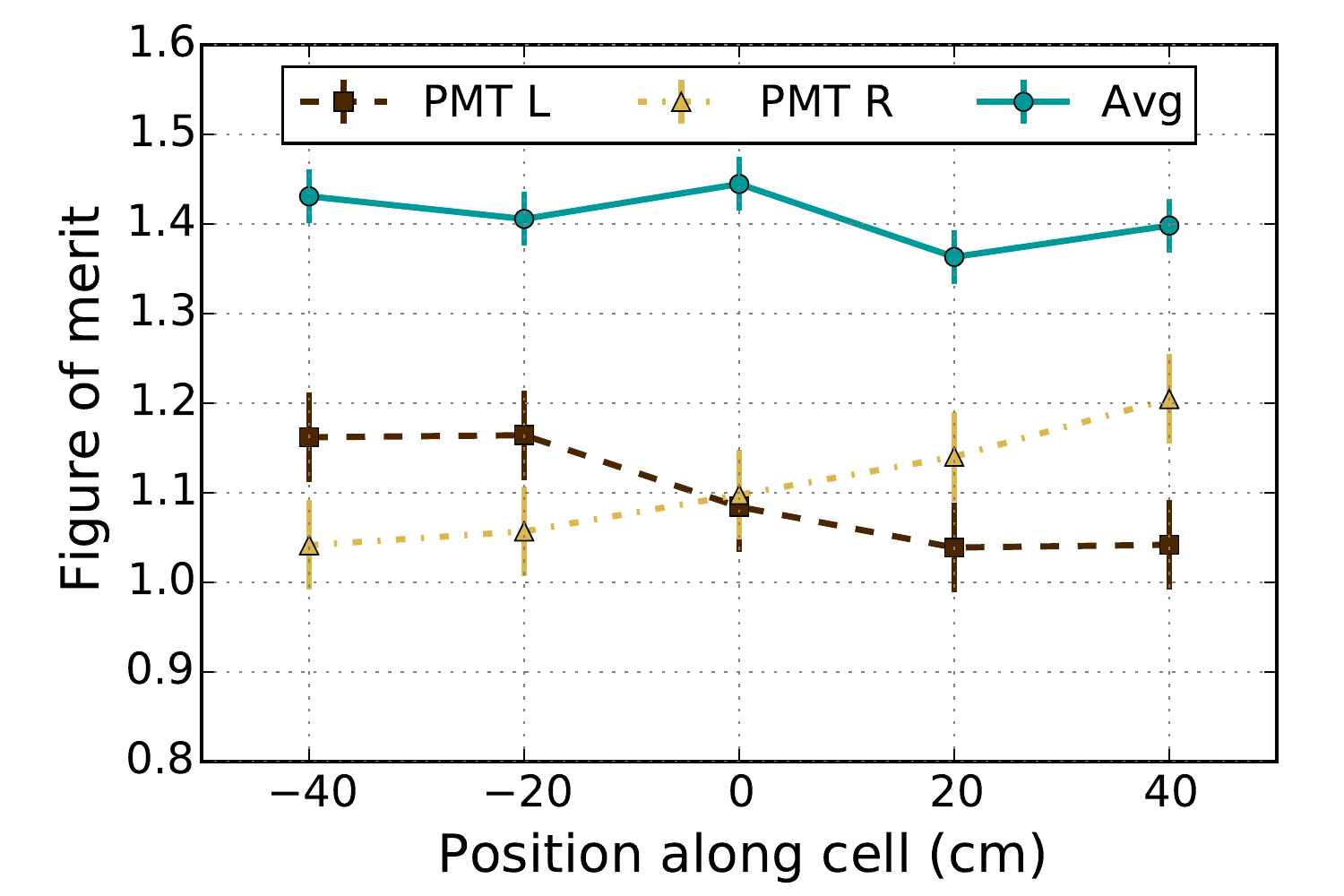}
\caption{
Left: PSD versus energy distribution from the summed L and R PMT signals in response to center-deployed \Cf source deployment.
PSD FOM values characterizing the level of neutron-gamma separation can be calculated for individual vertical energy slices of this distribution.
Right: Position variation in PSD FOM for the individual PMTs and their summed combination. }
\label{fig:singleVsTwoPMTPSD}
\end{center}
\end{figure}

As described in Section~\ref{sec:Data}, gamma- and neutron-like energy depositions in the cell can be distinguished by determining the fractional contribution of a waveform's long-time tail.
For all external-reflector configurations, an integration window from 10~ns before to 400~ns after the half-height of the waveform's leading edge was utilized, with the tail defined as all charge 28~ns after the leading edge half-height.
Figure~\ref{fig:singleVsTwoPMTPSD} shows the summed PMT energy versus PMT-averaged PSD value for the default configuration in response to a cell-midpoint ($z=0$~cm) \Cf deployment.
Two bands are clearly visible, with the lower band the result of electromagnetic interactions, and the upper band the result of interactions of heavy charged particles.
Peak and FWHM values used in calculating figures of merit were obtained by fitting two Gaussians to the PSD distributions in Figure~\ref{fig:singleVsTwoPMTPSD} for specific energy ranges, as described in Section~\ref{sec:Data}.
Statistical uncertainties on these FOM values are negligible, while fitting systematics from non-Gaussian tails produce $\sim$5\% variations in reported FOM from run to run.
As an example, from 1-3~MeV, an energy region associated with reactor antineutrino inverse beta decay positrons, the PSD FOM for the default configuration is 1.37.

Using the position reconstruction discussed in Sec.~\ref{sec:lightCollectionZdep}, it is possible to isolate the interaction locations of neutron events from a uniformly illuminating \Cf source.
The PSD FOM value from each PMT can then be calculated as a function of interaction position, as pictured in Figure~\ref{fig:singleVsTwoPMTPSD}.
The PMT-averaged PSD FOM is clearly superior to either individual PMT.
In addition, the FOM varies with position for individual PMTs, while the averaged FOM is uniform over the cell's length.

\section{Tests of Varied Cell Configurations}
\label{sec:designOpts}
To study the detector's performance in different configurations, three cell characteristics were varied: reflector type (specular or diffuse), un-coupled versus coupled reflectors, and single- verses double-ended PMT readout.
These variations allowed for identification of an optimal reflector and PMT configuration option, discussed in Section~\ref{sec:Interior}, when producing a cell configuration more closely matching the design parameters of a PROSPECT scintillator cell.

\subsection{Reflector Type}

\begin{figure}[htbp]
   \centering
   \includegraphics[width=0.48\textwidth]{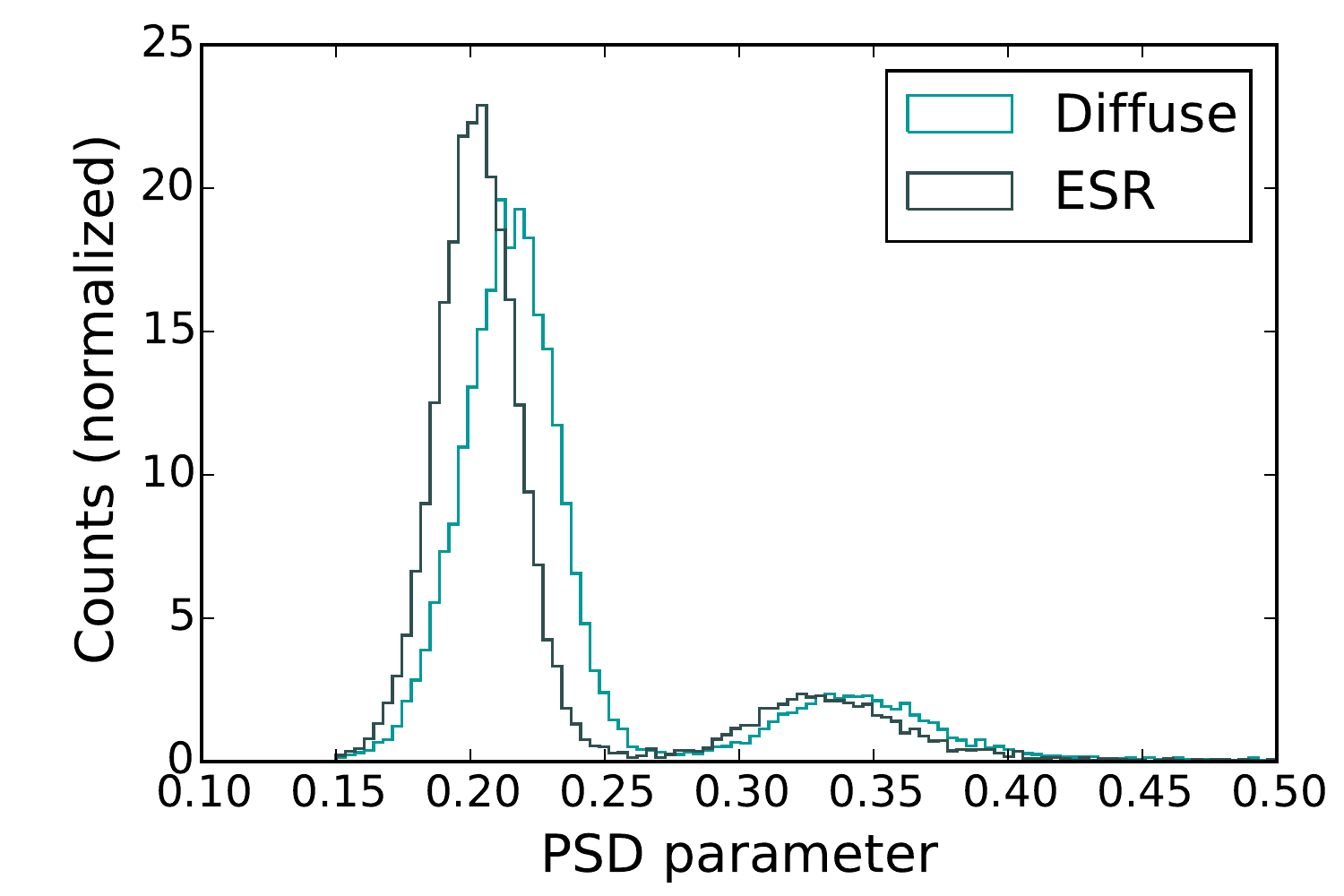}  
   \includegraphics[width=0.48\textwidth]{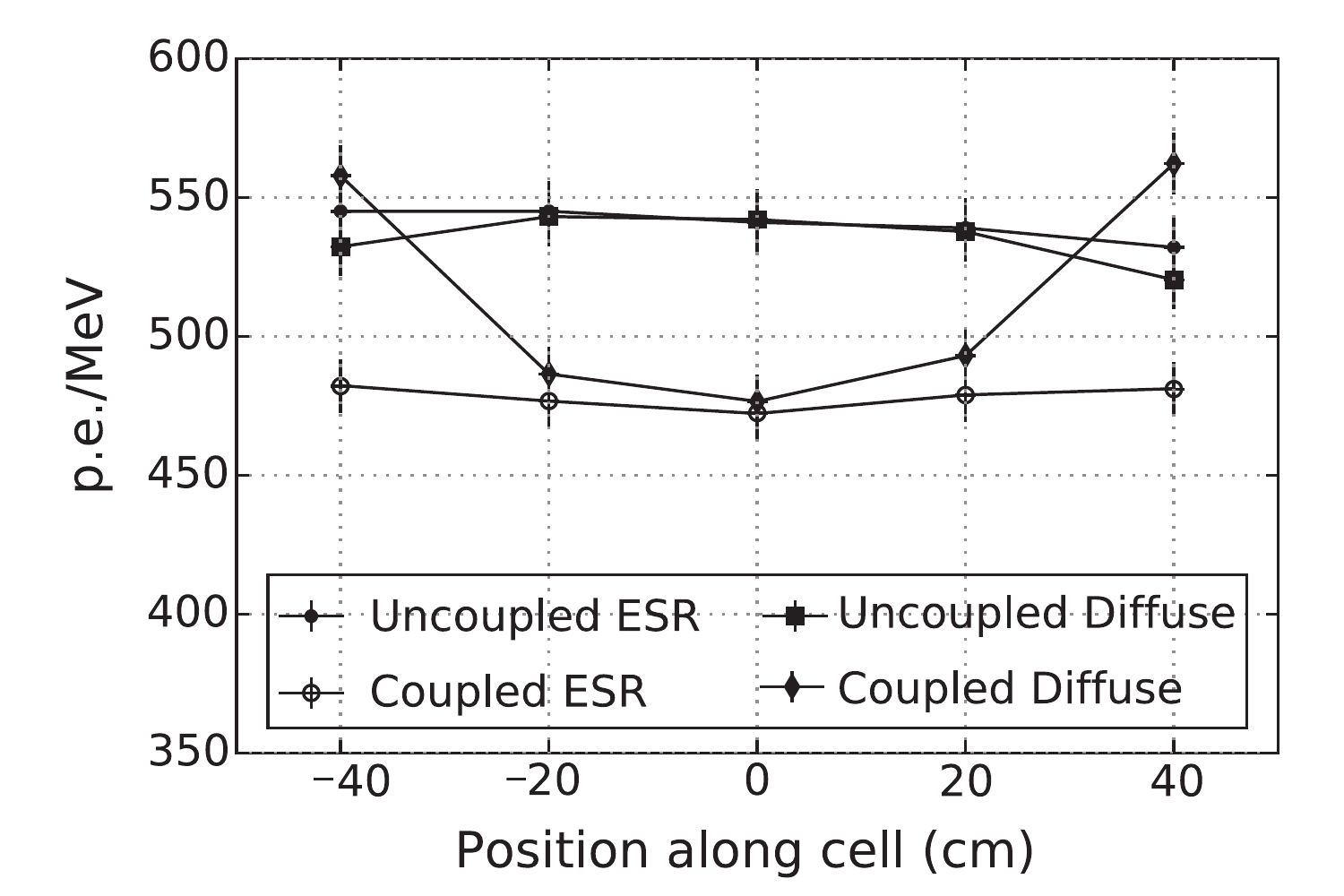}  
   \caption{
   Left: PSD distributions for uncoupled specular and diffuse reflector panels for events with energies between 1.0 and 3.0~MeV. 
   The diffuse panels produce broader neutron and gamma bands, degrading the overall PSD performance; this trend is exacerbated when TIR is reduced via reflector coupling.
Right: Light collection from the \Bi 0.85~MeV peak versus source position for coupled and un-coupled reflector panels (details in text). 
}
   \label{fig:specVsDiffusePSD}
\end{figure}

The total light collection efficiency and uniformity were measured for the various reflector types described in Section~\ref{sec:detector} using a \Bi source placed at multiple positions along the cell.  
With no external reflectors, total light collection is reduced by 33\% compared to the default configuration, indicating that a majority of the collected light propagates via total internal reflection.
When an air gap is maintained between an external reflector and the cell (`uncoupled'), light collection efficiencies remain within 3\% of the default configuration for both specular and diffuse reflectors.
This is in large part due to the large fraction of light that is propagates via TIR alone.
As shown in Figure~\ref{fig:specVsDiffusePSD}, PE yield uniformity along the cell for both of these data sets is also maintained.

The diffuse reflector panels have degraded PSD performance compared to the specular reflectors, even at the same light collection efficiency.
For events with energies between 1.0 and 3.0~MeV, the FOM is reduced from 1.37 in the default configuration to 1.16 when diffuse reflectors are used.
These tests demonstrate that for elongated cells, specular reflectors provide superior PSD compared to diffuse panels, even when the light collection is comparable.
This is likely due to the increased randomization of photon trajectories during travel in the cell and the subsequent effective broadening and lengthening of the photon arrival time distribution.
This effect is visible in Figure~\ref{fig:specVsDiffusePSD}, which shows noticeably larger mean late light fractions for both gamma- and neutron-like depositions.

\subsection{Coupled Versus Un-coupled Reflectors}

To study the impact of total internal reflection on the optical performance of the test detector, reflector panels were coupled with glycerin to two of the exterior surfaces on the opposite side of the cell. 
By removing TIR from two sides, a significant amount of light is no longer propagated via total internal reflection, and must instead rely on the character of the chosen reflector.
PE yield results of \Bi z-scans for un-coupled and two-glycerin-coupled data sets are shown in Figure~\ref{fig:specVsDiffusePSD}. 
The coupling of the reflector panels leads to a 14\% decrease in the total light collection, even for the high reflectivity specular ESR panels.
However, while the glycerin-coupled specular reflector tests demonstrate uniformity in light collection along the cell within the 2\% relative uncertainty, the coupled diffuse reflector data show a large non-uniformity of 15\%.
While non-uniformity in light collection may be compensated for using double-ended readout, there is also a degradation in PSD performance, with 1.0-3.0~MeV FOM values of 1.27 and 0.97 for specular and diffuse reflectors, respectively.  
Coupling of all four reflectors is likely to further reduce overall light yield and increase non-uniformity for the specular and diffuse reflector cases, respectively.

The importance of total internal reflection from interfaces between reflector and scintillator is thus clearly indicated by the data.
The air gaps of the un-coupled reflector cases provide the highest light collection and the best uniformity while PSD performance is optimized with air-coupled specular reflectors. 
We also note that, apart from air gaps, total internal reflection is also likely to be introduced by coating chosen reflector materials with materials exhibiting low refractive indices, a feature that will be further discussed in~Section~\ref{sec:Interior}.

\subsection{Single- vs Double-ended Readout}

\begin{figure}[h]
   \centering
   \includegraphics[width=0.48\textwidth]{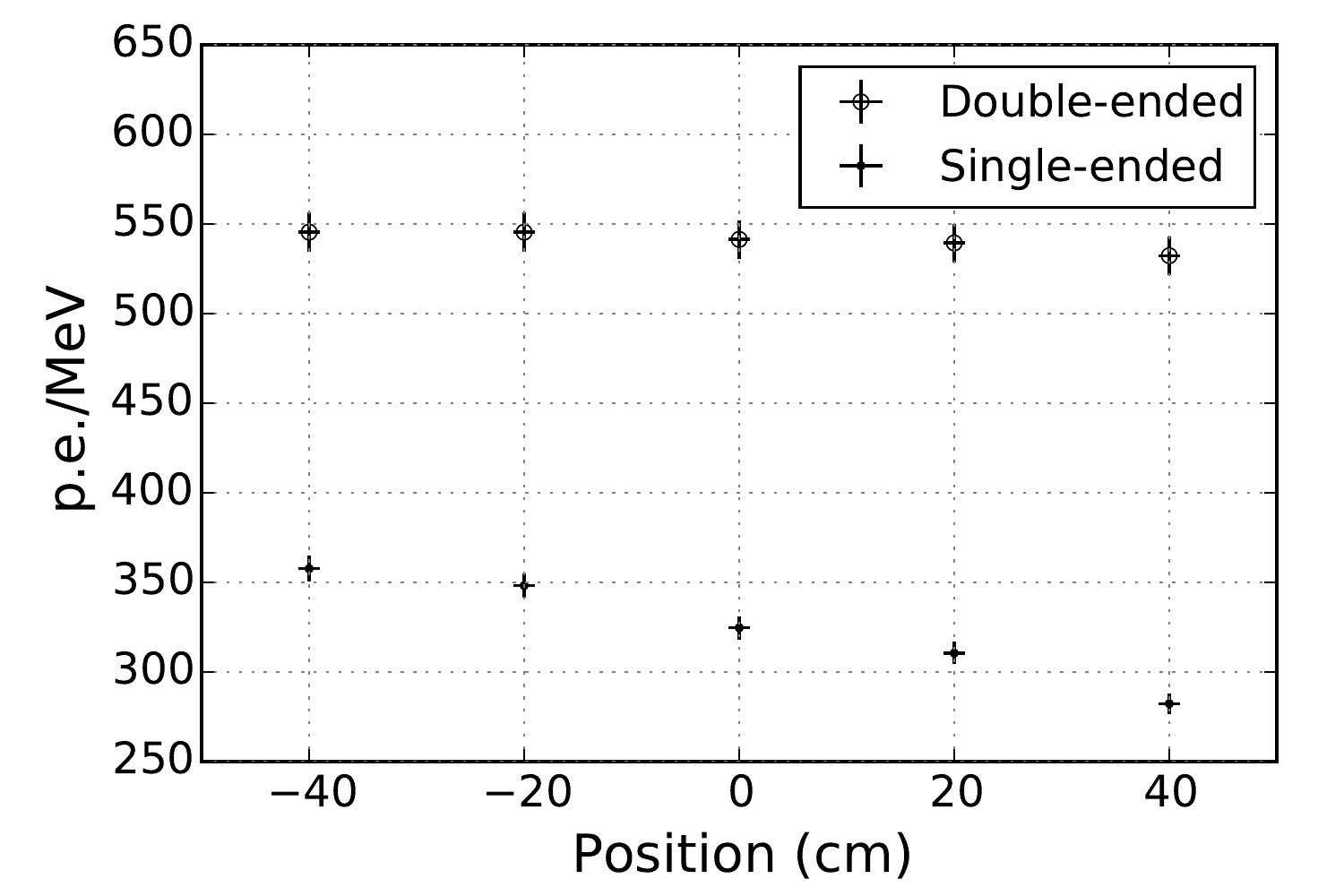}~ 
   \includegraphics[width=0.48\textwidth]{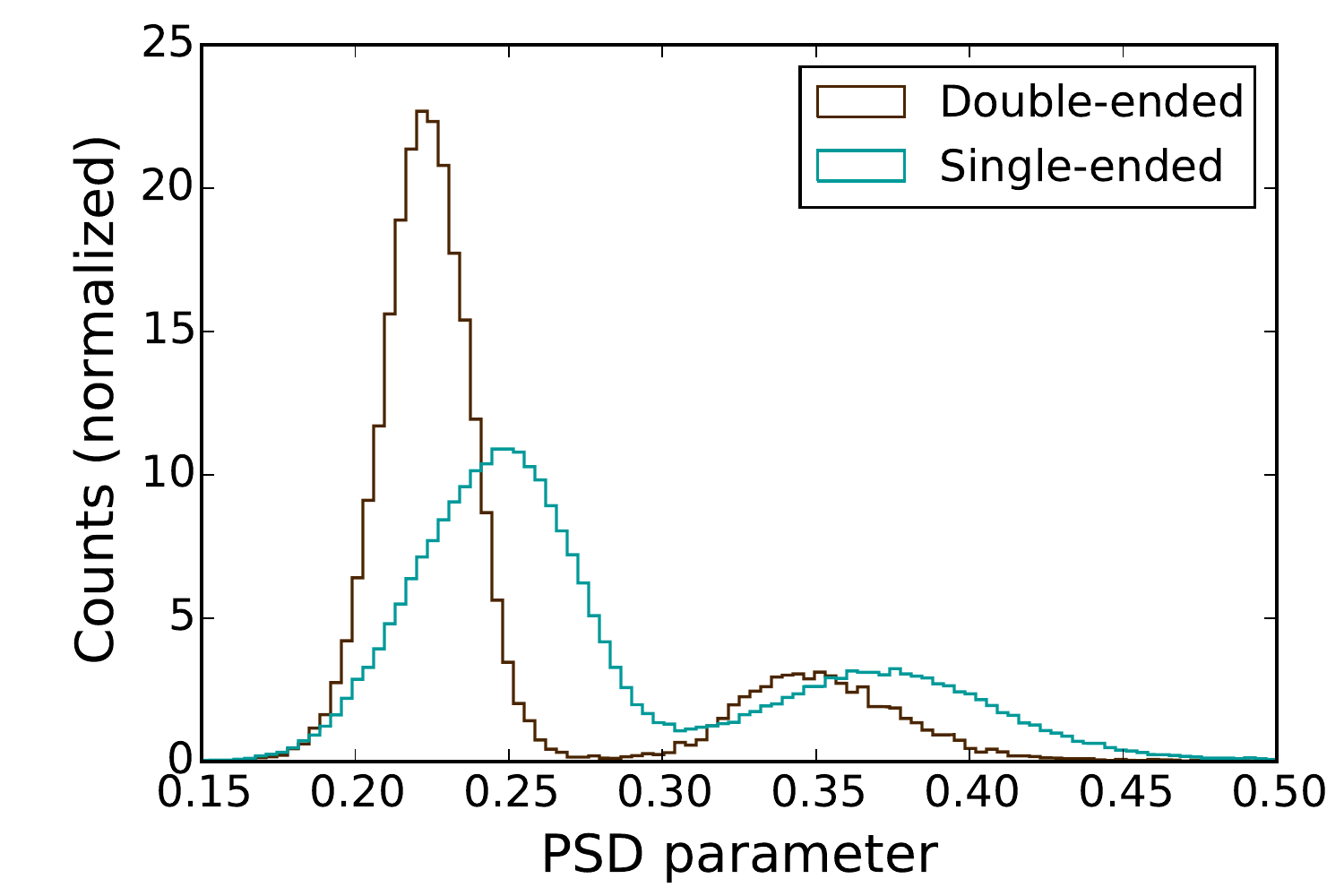}~ 
   \caption{Left: The light collection efficiency for single- and double-ended readout of the test detector when exposed to a \Bi source at multiple positions along the cell. 
   Right: A comparison of the [1.0-3.0]~MeV PSD performance from a centered \Cf source in single- and double-ended readout.  
   We note that the PSD peaks in these curves are slightly shifted with respect to the PSD distributions shown in Figure~7 due to replacement of an electronics cable and subsequent re-optimizing of PSD FOM.  
   This change had no net effect on PSD FOM for the default configuration.}
   \label{fig:singleVsDoublePMT}
\end{figure}

Explicit tests were also done with single-ended readout of the otherwise default test detector configuration to determine its suitability for elongated scintillator cell design.
One PMT was removed from the default configuration and replaced with a single sheet of air-coupled Solar Mirror 2020 specular reflector.  
As shown in Figure~\ref{fig:singleVsDoublePMT}, maximum PE yield for this configuration was 35\% lower than the default configuration for a \Bi center source deployment, with a total PE yield non-uniformity of 26\%.
This configuration thus provides both worse absolute collection efficiency and energy resolution than the default configuration, while also providing no method for position reconstruction or non-uniformity correction along the cell.
Figure~\ref{fig:singleVsDoublePMT} shows that PSD performance is also degraded when operating the detector in single-ended readout. 

\subsection{Summary of Variations}

\begin{figure}[h]
   \centering
   \includegraphics[width=0.8\textwidth]{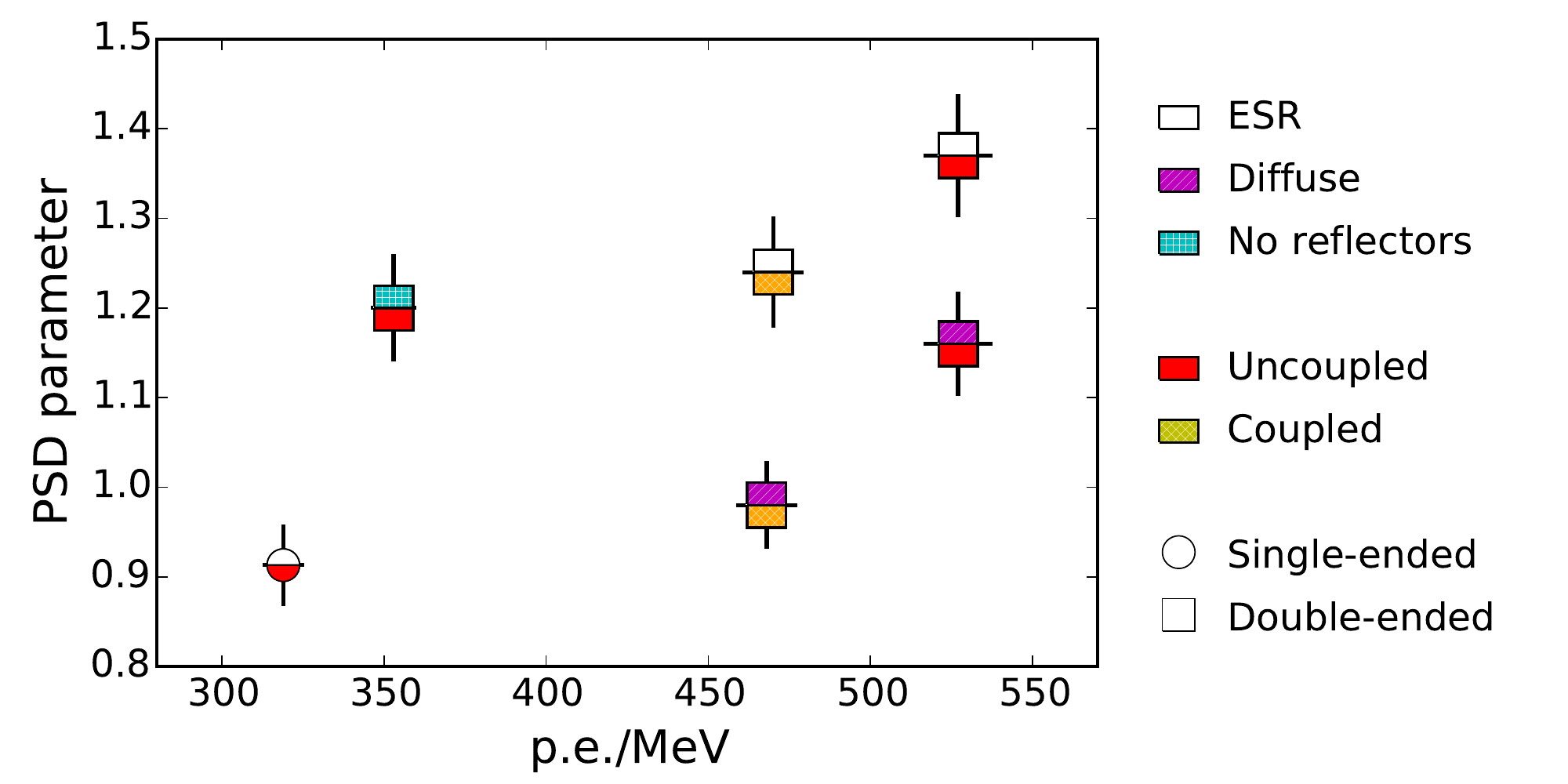} 
   \caption{Distribution of the PSD performance versus light collection for events between 1.0 and 3.0 MeV in response to a center \Cf source deployment for various setups described in this section.  The legend describes the different configuration options for each point.  
Error bars for PSD FOM and PE yield are defined as given in the text.
The spread in both quantities highlights the importance of carefully optimizing cell design to both light collection and PSD.}
   \label{fig:PSDvsPE}
\end{figure}

Figure~\ref{fig:PSDvsPE} shows the spread of PSD FOMs versus light collection (PE/MeV) for each setup discussed in this work. 
While there is some correlation between light collection and PSD performance, it is clearly necessary to optimize both parameters for a given geometry. 
Given the appropriate optimization, elongated scintillator cell geometries are capable of delivering excellent energy resolution and background rejection capabilities.

\section{Optimized Test Cell Configuration and Results}
\label{sec:Interior}

Having determined the performance of relevant configurations for the test detector with easily-adjustable external reflectors, we now turn to the design and characterization of an optimized configuration that more closely mirrors the intended PROSPECT cell design.
In particular, by deploying cell-internal reflectors, the test detector can more faithfully represent the actual optical and mechanical characteristics of the baseline PROSPECT cells in several key aspects:
\begin{itemize}
  \item{Internal reflectors imply a direct interface between reflector and scintillator.  The PROSPECT energy resolution physics requirement demands a low inactive mass fraction and close-packed cells, necessitating direct contact of reflector panel and scintillator.}
  \item{The test detector only reproduces the baseline PROSPECT optical cell cross-section when internal reflectors are installed.  
  The presence of the 1~cm thick acrylic cell walls inside the exterior reflectors inflates the effective optical cell dimension to 17.2~cm, producing a large amount of light loss due to the PMT-cell cross-section mis-matching.  
  Internal reflectors reduce this light loss while yielding a $\sim$14.5~cm cell cross-section, chosen to match the total 12.5~cm PMT diameter plus 2 additional~cm needed for the structural support of the PMTs and target segmentation system.}
  \item{Internal reflectors can extend beneath the 2.5~cm flange on each cell end, unlike external reflectors, reducing light loss from these un-covered regions, which are not present in the baseline PROSPECT cell.}
\end{itemize}

The studies in the previous section clearly indicate that PSD, PE collection, and uniformity are optimized by utilizing specular reflectors and by exploiting TIR to the greatest extent possible.
For these reasons, we chose to produce internal reflectors utilizing ESR specular reflector with a chemically inert, and scintillator compatible, FEP Teflon cover.
Additionally, FEP has an index of refraction (n$\approx$1.32) significantly different from the scintillator (n$\approx$1.56), maximizing TIR in the absence of an air gap.

Four internal reflector panels were produced by laminating ESR specular reflector to one side of four 3~mm thick acrylic panels with 45 $\deg$-bevelled long edges whose length and width matched that of the full cell interior.  
Oversized FEP sheets of 51~micron thickness were then laminated around each panel and heat-bonded around the panel edges, producing a leak-tight FEP casing around the central panel and ESR.
With the cell emptied and end flanges removed, the four individually-encased panels were then held together to produce a tube of square cross-section that was then slid with minimal clearance into the cell.
Panels were installed such that excess FEP bond edging was tucked behind adjacent panels.
Thinner UVT endplates of 1.2~cm thickness were then attached to further reduce the total non-reflective length between the cell and each PMT.
The detector was then re-filled, with PMTs and electronics identical to previous configurations.

\begin{figure}[h]
   \centering
   \includegraphics[width=0.48\textwidth]{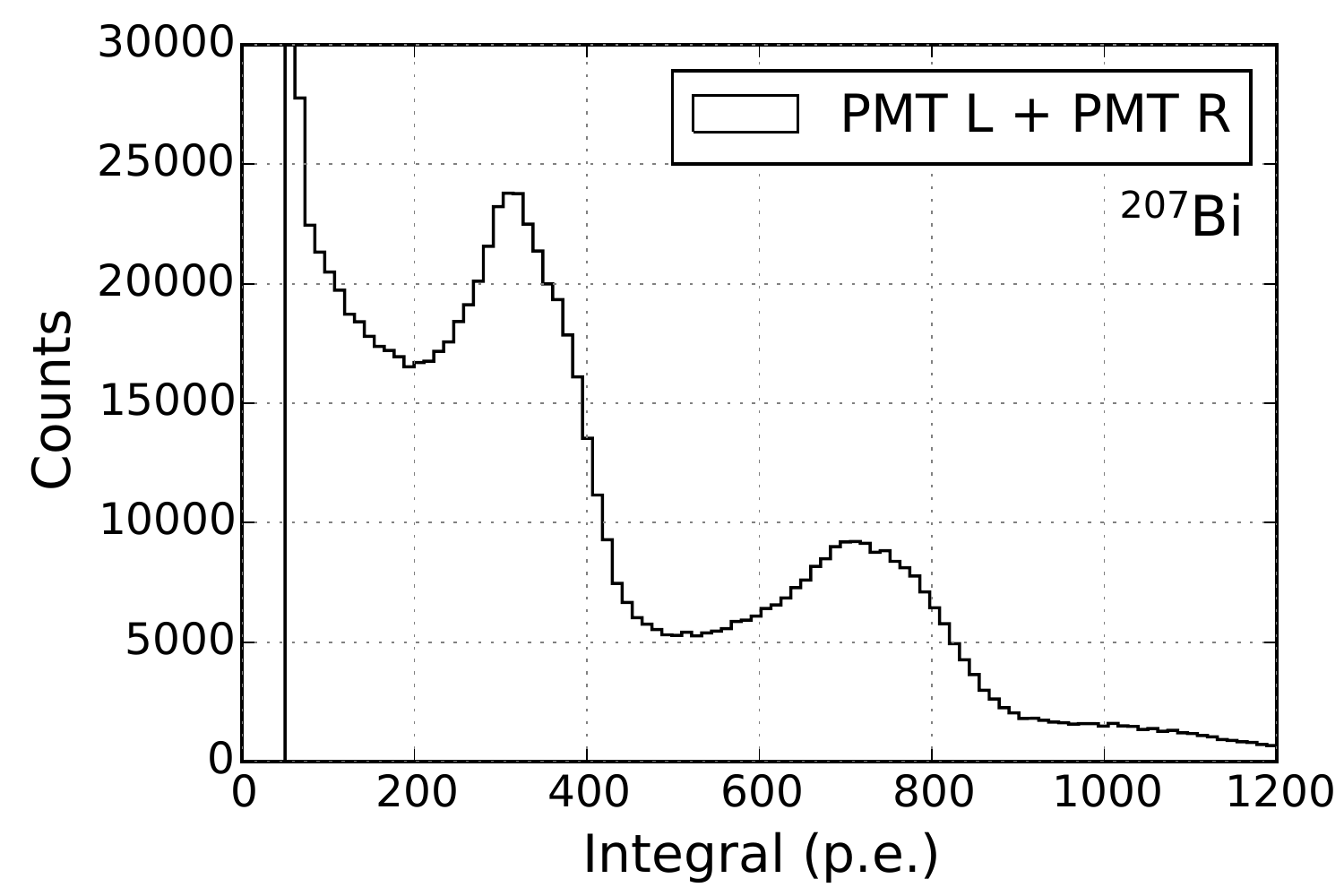}
   \includegraphics[trim=1.3cm 0.1cm 1.3cm 0.2cm, clip=true, width=0.48\textwidth]{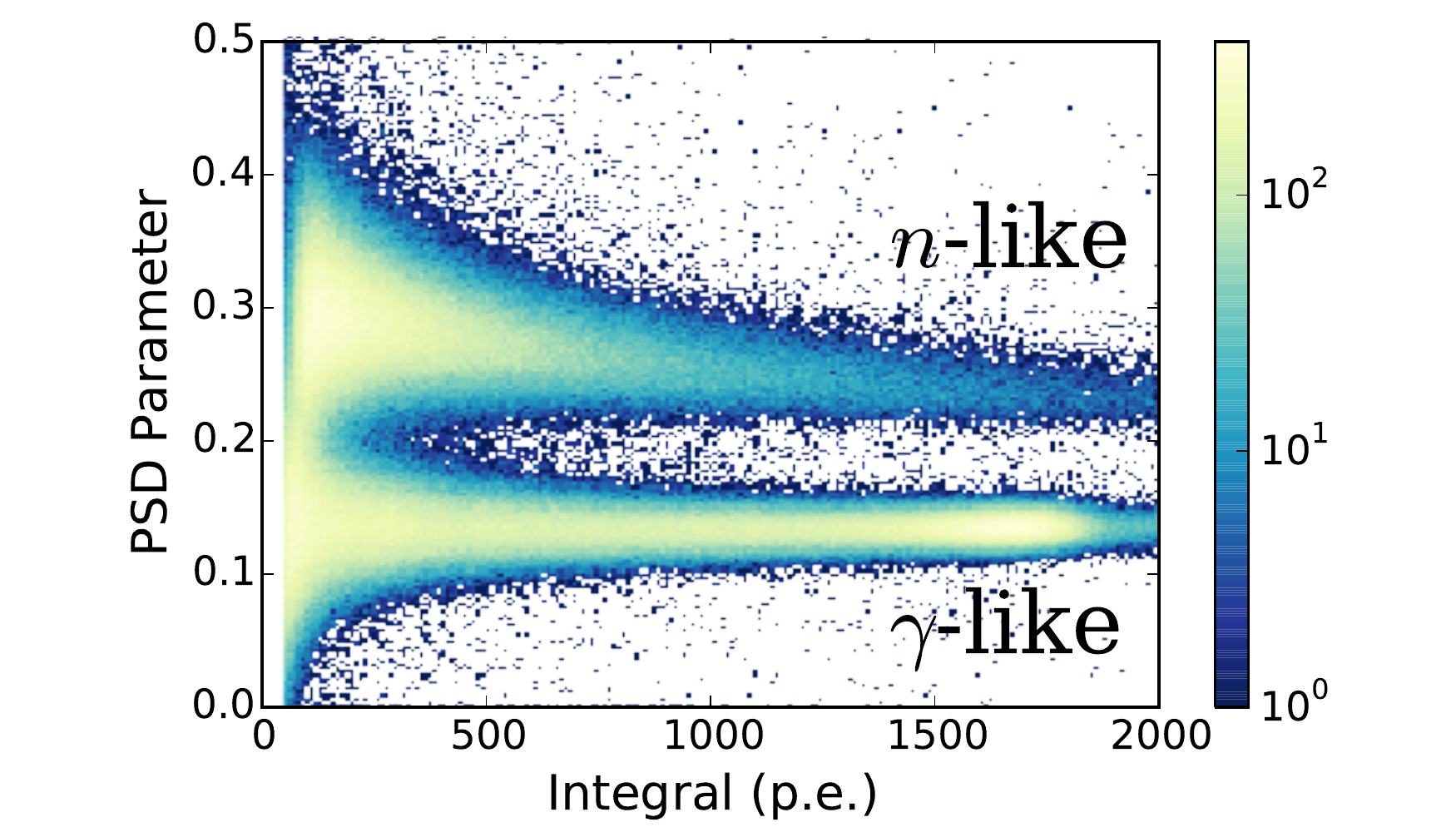}
   \caption{Left: Summed PE spectrum from both PMTs in the `internal reflector' configuration in response to a cell-center \Bi deployment.
     Right: PSD parameter versus summed PE for the internal reflector configuration in response to a cell-center \Cf-deployment.
}
   \label{fig:InternalSource}
\end{figure}

The performance of the `internal reflector' configuration was then characterized using the \Bi and \Cf sources.  
The summed PE spectrum of the \Bi source deployed at cell center can be seen in Figure~\ref{fig:InternalSource}.  
Total averaged PE yield as determined by the same three \Bi and \Cf spectrum peaks is 841~$\pm$~17~PE/MeV, 58\% higher than that of the `default' external reflector configuration.  
Z-scans with \Bi once again show PE yields consistent along the cell to within the 2\% systematic uncertainty.
PMT-summed PE and averaged PSD values for the center-deployed \Cf source, also shown in Figure~\ref{fig:InternalSource}, exhibit very clearly defined gamma and neutron bands.
PMT-averaged PSD FOM for \Cf-produced triggers in the 1.0-3.0~MeV region improves from a maximum of 1.37 for any external reflector configuration to 1.73 for the internal reflector configuration.

\begin{figure}[!h]
\begin{center}
\includegraphics[width=0.55\textwidth]{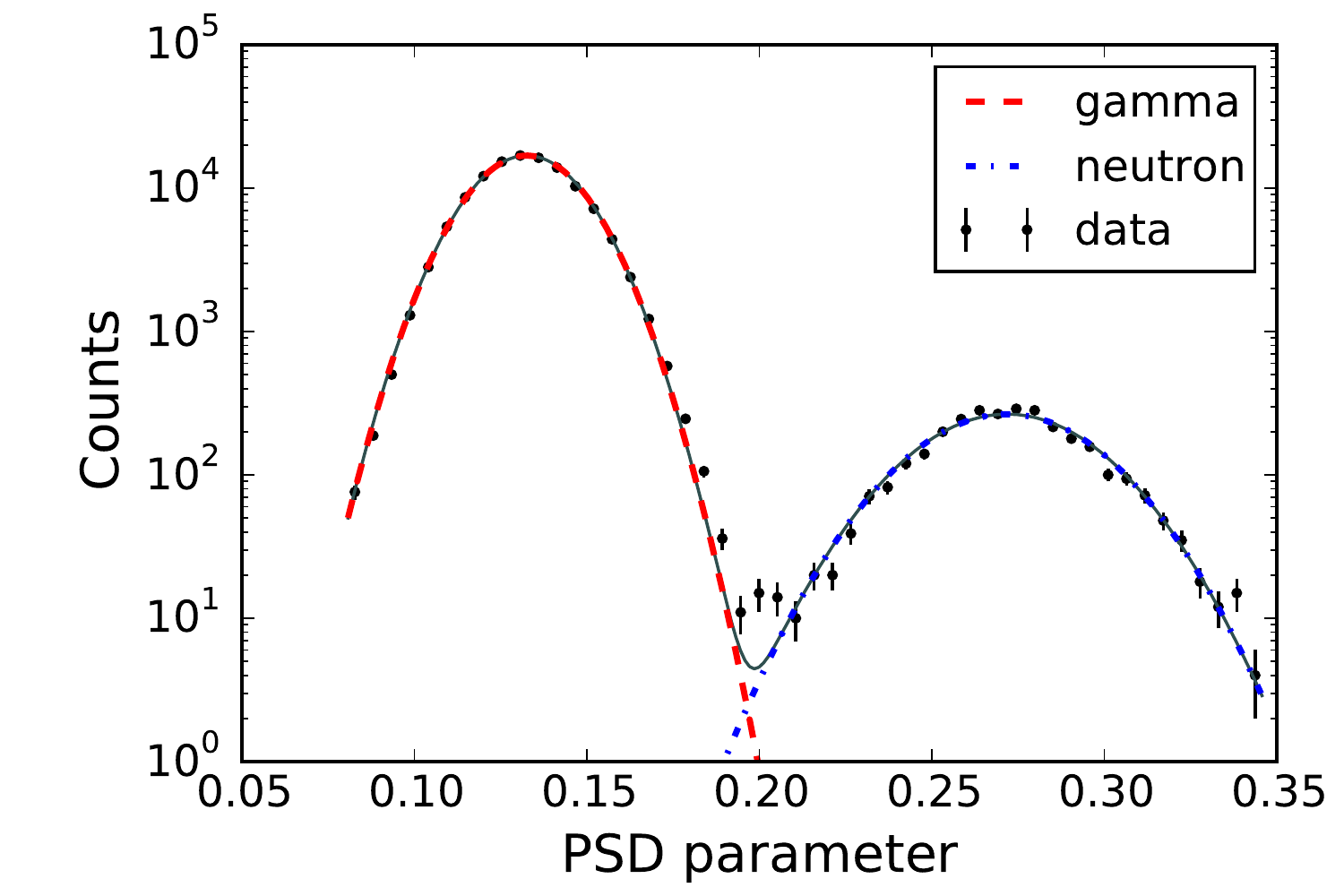}
\caption{PSD distribution of ambient background events in the [0.5-0.7]~MeV energy range related to lithium neutron capture in organic liquid scintillator. Excellent separation between neutron and gamma events is preserved, despite the low number of neutron events.}
\label{fig:backgroundPSD}
\end{center}
\end{figure}

To demonstrate the background rejection ability of this test detector configuration, it was operated for one hour in the ambient background environment.
The PSD distribution for ambient events in the [0.5-0.7]~MeV energy band is shown in Figure~\ref{fig:backgroundPSD}, where clear separation between light and heavy charged particle bands is observed.
Using the fitted two Gaussian curve, and assuming true Gaussian tails, a cut in PSD parameter below 0.18 will reduce the gamma-ray contribution by more than four orders of magnitude while preserving more than 99.9\% of heavy charged particle events.

\section{Conclusions and Outlook}
\label{sec:conc}

The optical response of a meter-long liquid scintillator cell filled with 23~liters of EJ-309 scintillator has been characterized utilizing gamma and spontaneous fission calibration sources.
The results are useful for benchmarking the expected response of individual scintillator cells in the PROSPECT antineutrino detector, while also providing more general guidance to the construction of elongated liquid scintillator detectors.

For the elongated cell geometry and reflector configuration most closely resembling the baseline PROSPECT cell design, a light collection efficiency of 841$\pm$17 PE/MeV is obtained.
Excellent pulse shape discrimination is achieved in this geometry, demonstrating that a large volume detector can provide the particle identification capabilities required for near-reactor on-surface antineutrino detectors. 
PSD in the best-performing configuration has been demonstrated to yield over four orders of magnitude rejection of electromagnetic interactions in the low-energy region of [0.5-0.7]~MeV where PROSPECT will detect neutron captures on $^6$Li. 
A comparable rejection power is observed in the inverse beta decay positron energy region. 
Both pulse shape discrimination and light collection have been shown to be highly uniform along this cell configuration. 
Timing-based and charge-based position reconstruction in such a double-ended-readout cell has been demonstrated to localize events within 5~cm along the cell.
These features are of significant value in producing a high-resolution, low-background short-baseline reactor antineutrino measurement.

We have also produced systematic studies in three key areas of design crucial for elongated scintillator cells.  
First, specular reflectors have been shown to be superior to diffuse reflectors in PSD performance, even when producing comparable PE yields.
In addition, when diffuse reflection is the dominant propagation method, light collection non-uniformities are significantly increased.
Of equal importance, our studies show that un-coupled diffuse or specular reflectors preserving total internal reflections along the cell provide higher light collection and better PSD performance than coupled reflectors.
In the case of the tests described here, air coupling between reflector and scintillator could be achieved easily by placing reflector panels on the exterior of the cell.
For close-packed arrays of elongated scintillator cells, such as would be deployed in a high-resolution reactor antineutrino detector, low-index reflector coverings provide a very effective replacement for such an air gap, as demonstrated with the test detector internal reflector configurations.
Finally, double-ended readout has been shown to yield enhanced pulse shape performance and light collection efficiency and uniformity.
These response gains are in addition to the valuable position reconstruction benefits that come from a two-PMT configuration.

\section{Acknowledgements}
This material is based upon work supported by the U.S. Department of Energy Office of Science and the National Science Foundation. 
Additional support for this work is provided by Yale University and the Illinois Institute of Technology. 
We gratefully acknowledge the support and hospitality of the High Flux Isotope Reactor, managed by UT-Battelle for the U.S. Department of Energy. 

\bibliographystyle{h-physrev}
\bibliography{PROSPECT20}

\end{document}